\documentclass[12pt]{article}
\usepackage{a4wide,amsmath,graphicx,multirow}
\numberwithin{equation}{section}

\begin{document}

\renewcommand{\thefootnote}{\fnsymbol{footnote}}
\begin{flushright}
TTP11-09
\end{flushright}
\begin{center}
\LARGE Integration by parts\\[0.5em]
\Large An introduction%
\footnote{Extended version of the lectures
at the school ``Computer algebra and particle physics 2011'',
DESY Zeuthen, March 21--25, 2011.}\\[1.5em]
\large A.\,G.~Grozin%
\footnote{Permanent address: Budker Institute of Nuclear Physics, Novosibirsk}\\[0.5em]
\large Institut f\"ur Theoretische Teilchenphysik,\\
Karlsruher Institut f\"ur Technologie
\end{center}

\begin{abstract}
Integration by parts is used to reduce scalar Feynman integrals to master integrals.
\end{abstract}

\renewcommand{\thefootnote}{\arabic{footnote}}
\setcounter{footnote}{0}

\section{Introduction}
\label{S:0}

Solution of any problem in perturbative quantum field theory
includes several steps:
\begin{enumerate}
\item diagrams generation, classification into topologies, routing momenta;
\item tensor and Dirac algebra in numerators, reduction to scalar Feynman integrals;
\item reduction of scalar Feynman integrals to master integrals;
\item calculation of master integrals.
\end{enumerate}
For a sufficiently complicated problem, all of them must be completely automated.
The number of Feynman diagram in a problem can be very large.
They can be classified into a moderate number of generic topologies.
In each of them, a large number of scalar Feynman integrals is required
(especially if expansion in some small parameter up to a high degree is involved).
They are not independent: there are integration-by-parts (IBP) recurrence relations.
These relations can be used to reduce all these scalar Feynman integrals
to a small number of master integrals~\cite{CT:81}
(see also textbooks~\cite{FIC,G:07}).
The number of master integrals can be proved to be finite~\cite{SP:10},
but the proof is not constructive ---
it provides no method of reduction.

IBP relations are sometimes used together
with recurrence relations in $d$~\cite{T:96,L:10};
we don't discuss these relations here.
Several methods of calculation of master integrals
also require the reduction problem to be solved:
differential equations~\cite{K:91,R:97}
(see also Chapter~7 in~\cite{FIC} and the review~\cite{AM:07}),
recurrence relations in $d$~\cite{L:10},
gluing~\cite{BC:10}.
Other methods, e.\,g., Mellin--Barnes representation, don't depend on IBP.
Here we shall discuss only reduction (step 3);
methods of calculation of master integrals is a separate (and large) topic.

If the problem involves small ratios of external momenta and masses,
there is an additional step --- expansion in these ratios
using the method of regions~\cite{S:02}.
It can produce new kinds of denominators in Feynman integrals.
It is done after (or before) the step 2.

\section{Feynman graphs and Feynman integrals}
\label{S:Intro}

Let's consider a diagram with external momenta $p_1$, \dots, $p_E$.
If we are considering a generic kinematic configuration,
this means that the diagram has $E+1$ external leg
(it can have a larger number of legs,
if we are interested in a restricted kinematics
with linearly dependent momenta of these legs).
An $L$-loop diagram has $L$ loop (integration) momenta $k_1$, \dots, $k_L$.
Let $q_i = k_1$, \dots, $k_L$, $p_1$, \dots, $p_E$ ($i\in[1,M]$, $M=L+E$)
be all the momenta.
The diagram has $I$ internal lines with the momenta $l_1$, \dots, $l_I$;
they are linear combinations of $q_i$.
Let $s_{ij}=q_i\cdot q_j$ ($j\ge i$);
$N_E=E(E+1)/2$ scalar products $s_{ij}$ with $i>L$ are external kinematic quantities, and
\begin{equation}
N = \frac{L(L+1)}{2} + LE
\label{Intro:N}
\end{equation}
scalar products with $i\le L$ are integration variables.

Massless and massive lines in a Feynman graph for a scalar Feynman integral are
\begin{equation}
\raisebox{-6.3mm}{\begin{picture}(22,9)
\put(11,6){\makebox(0,0){\includegraphics{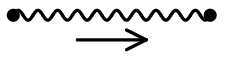}}}
\put(11,1.5){\makebox(0,0){$l$}}
\end{picture}} = \frac{1}{- l^2 - i0}\,,\qquad
\raisebox{-3.3mm}{\begin{picture}(22,6)
\put(11,4.5){\makebox(0,0){\includegraphics{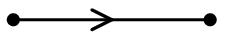}}}
\put(11,1.5){\makebox(0,0){$l$}}
\end{picture}} = \frac{1}{m^2 - l^2 - i0}\,.
\label{Intro:line}
\end{equation}
In HQET, the propagator
\begin{equation}
\raisebox{-3.3mm}{\begin{picture}(22,6)
\put(11,4.5){\makebox(0,0){\includegraphics{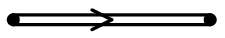}}}
\put(11,1.5){\makebox(0,0){$l$}}
\end{picture}} =
\frac{1}{-2 l \cdot v - i0}
\label{Intro:HQET}
\end{equation}
appears, where the heavy-particle velocity $v$ ($v^2=1$)
appears in the Lagrangian, see the textbooks~\cite{MW:00,G:04}.
Similar propagators appear in SCET, but with $v^2=0$.
In NRQED, NRQCD the denominator is more complicated:
$- 2M l\cdot v + (l\cdot v)^2 - l^2 - i0$.
Instead of using effective field theories,
one may follow a more diagrammatic approach
and expand the ordinary propagators in specific regions of $k_i$~\cite{S:02}.
In all cases, the denominator of a propagator
is quadratic or linear in the line momentum $l$.

\begin{figure}[ht]
\begin{center}
\begin{picture}(110,80)
\put(25,19.75){\makebox(0,0){\includegraphics{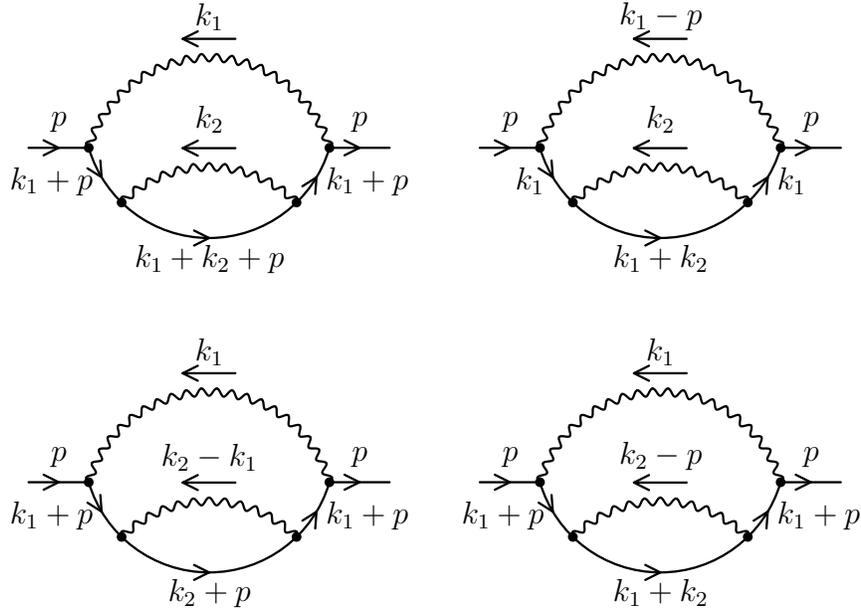}}}
\put(25,33.5){\makebox(0,0)[b]{$k_1$}}
\put(25,5){\makebox(0,0)[t]{$k_2+p$}}
\put(25,19.49){\makebox(0,0)[b]{$k_2-k_1$}}
\put(5,20){\makebox(0,0)[b]{$p$}}
\put(45,20){\makebox(0,0)[b]{$p$}}
\put(40.5,13.5){\makebox(0,0)[l]{$k_1+p$}}
\put(9.5,13.5){\makebox(0,0)[r]{$k_1+p$}}
\put(85,19.75){\makebox(0,0){\includegraphics{rout.eps}}}
\put(85,33.5){\makebox(0,0)[b]{$k_1$}}
\put(85,5){\makebox(0,0)[t]{$k_1+k_2$}}
\put(85,19.49){\makebox(0,0)[b]{$k_2-p$}}
\put(65,20){\makebox(0,0)[b]{$p$}}
\put(105,20){\makebox(0,0)[b]{$p$}}
\put(100.5,13.5){\makebox(0,0)[l]{$k_1+p$}}
\put(69.5,13.5){\makebox(0,0)[r]{$k_1+p$}}
\put(25,64.25){\makebox(0,0){\includegraphics{rout.eps}}}
\put(25,78){\makebox(0,0)[b]{$k_1$}}
\put(25,49.5){\makebox(0,0)[t]{$k_1+k_2+p$}}
\put(25,64.45){\makebox(0,0)[b]{$k_2$}}
\put(5,64.5){\makebox(0,0)[b]{$p$}}
\put(45,64.5){\makebox(0,0)[b]{$p$}}
\put(40.5,58){\makebox(0,0)[l]{$k_1+p$}}
\put(9.5,58){\makebox(0,0)[r]{$k_1+p$}}
\put(85,64.25){\makebox(0,0){\includegraphics{rout.eps}}}
\put(85,78){\makebox(0,0)[b]{$k_1-p$}}
\put(85,49.5){\makebox(0,0)[t]{$k_1+k_2$}}
\put(85,64.45){\makebox(0,0)[b]{$k_2$}}
\put(65,64.5){\makebox(0,0)[b]{$p$}}
\put(105,64.5){\makebox(0,0)[b]{$p$}}
\put(100.5,58){\makebox(0,0)[l]{$k_1$}}
\put(69.5,58){\makebox(0,0)[r]{$k_1$}}
\end{picture}
\end{center}
\caption{Momentum routings}
\label{F:Routing}
\end{figure}

The choice of the integration momenta $k_i$ is not unique.
Momentum conservation at each vertex is the only restriction.
Some examples of different momentum routings for a single diagram
are shown in Fig.~\ref{F:Routing}.
It is not easy for a program to recognize that two Feynman integrals
can be transformed into each other by linear substitutions of $k_i$.
In fact, the freedom of choice is much larger
than suggested by Fig.~\ref{F:Routing},
because all the momenta $p$, $k_1$, $k_2$ can flow along all lines
with some continuous weights.
The value of a given Feynman integral cannot depend on a choice
of momentum routing
(all choices agree with Feynman rules of the theory,
and are equally good).

\begin{figure}[ht]
\begin{center}
\begin{picture}(98,36)
\put(28,20){\makebox(0,0){\includegraphics{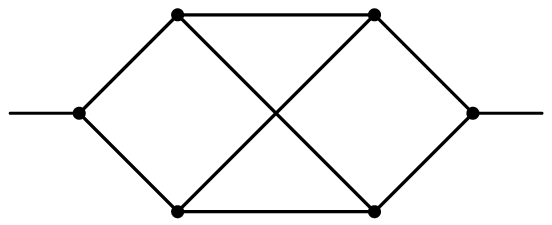}}}
\put(28,1.5){\makebox(0,0){a}}
\put(82,20){\makebox(0,0){\includegraphics{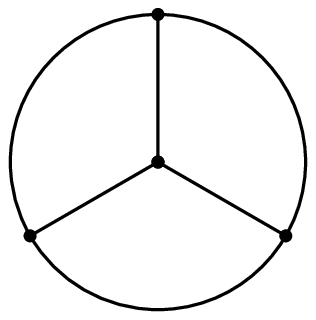}}}
\put(82,1.5){\makebox(0,0){b}}
\end{picture}
\end{center}
\caption{Symmetric diagrams}
\label{F:Sym}
\end{figure}

Many diagrams have symmetries.
For example, the non-planar self-energy diagram in Fig.~\ref{F:Sym}a
may be reflected in a horizontal mirror.
It also may be reflected in a vertical mirror;
this operation changes the external momentum $p\to-p$,
and hence, in order to get an identical integrand,
we have to do the substitution $k_i\to-k_i$ of the integration momenta.
One more symmetry is less obvious:
you can keep the 3 left vertices at rest,
and rotate the 3 right ones around the horizontal axis by $\pi$.
The diagram in Fig.~\ref{F:Sym}b (where all the lines have the same mass)
is very symmetrical.
It can be represented as a tetrahedron,
and thus has the full tetrahedron symmetry group:
rotations by $n\pi/3$ around 4 axes and reflections in 6 planes.
It is difficult for a program to find symmetries of a Feynman integral,
because symmetry transformations often have to be followed
by integration-momenta substitutions to write the transformed integral
in its original form.

We shall consider Feynman graphs containing only 3-legged vertices (generic topologies).
They have the maximum number of internal lines $I$.
All the remaining diagrams can be obtained from these ones
by shrinking some internal lines, i.\,e., raising some denominators to the power 0.
They belong to reduced topologies.
If the diagram we want contains a 4-legged vertex,
we can split this vertex into 2 3-legged ones with an internal line between them,
and say that this internal propagator is shrunk;
thus, our diagram becomes a particular case of a generic one
with 0 power of some internal line.
This generalization is not unique:
a diagram belonging to a reduced topology can be obtained from several different
generic topologies by shrinking some internal lines.

For $(E+1)$-legged tree diagrams, the number of internal lines is $I=E-2$
(just 1 vertex has $E=2$ and $I=0$; adding an external leg increases $I$ by 1).
Adding a loop (i.\,e.\ connecting 2 points on some lines) increases $I$ by 3.
So, $I$ for $(E+1)$-legged $L$-loop diagrams is
\begin{equation}
I = 3L + E - 2\,,\qquad
N - I = \frac{(L-1)(L+2E-4)}{2}\,.
\label{Intro:legged}
\end{equation}
Therefore, if $L\ge\max(2,5-2E)$,
then there are more scalar products than denominators of propagators.
In such a case, it is not possible to express all $s_{ij}$
as linear combinations of the denominators.
For self-energy diagrams ($E=1$), this happens starting from $L=3$ loops;
for diagrams with more legs --- starting from $L=2$ loops.

Vacuum diagrams ($E=0$) have to be considered separately.
The simplest diagram with 2 3-legged vertices has $L=2$ and $I=3$.
Adding a loop increases $I$ by 3, and hence
\begin{equation}
I = 3 (L-1)\,,\qquad
N-I = \frac{(L-2)(L-3)}{2}\,.
\label{Intro:vac}
\end{equation}
Scalar products which cannot be expressed via denominators
appear starting from $L=4$ loops.

We want all scalar products $s_{ij}$ with $i\le L$ to be expressible
as linear functions of the denominators $D_a$.
Therefore, we add irreducible numerators (linear functions of $s_{ij}$)
$D_{I+1}$, \dots, $D_N$:
\begin{equation}
D_a = \sum_{i=1}^L \sum_{j=i}^M A_a^{ij} s_{ij} + m^2_a\,.
\label{Intro:Ds}
\end{equation}
The only requirement is that the full set $D_1$, \dots, $D_N$ is linearly independent.
Then we can solve for $s_{ij}$:
\begin{equation}
s_{ij} = \sum_{a=1}^N A^a_{ij} (D_a - m^2_a)\,.
\label{Intro:sD}
\end{equation}
If we view the pair $(ij)$ with $i\in[1,L]$ and $j\ge i$ as a single index
(it has $N$ different values),
then the matrix $A^a_{ij}$ is the inverse of $A_a^{ij}$.

Now we define the scalar Feynman integral
\begin{equation}
\begin{split}
&I(n_1,\ldots,n_N) = \frac{1}{(i\pi^{d/2})^L}
\int d^d k_1\cdots d^d k_L\,f(k_1,\ldots,k_L,p_1,\ldots,p_E)\,,\\
&f(k_1,\ldots,k_L,p_1,\ldots,p_E) = \frac{1}{D_1^{n_1}\cdots D_N^{n_N}}\,.
\end{split}
\label{Intro:I}
\end{equation}
Of course, for irreducible numerators $n_a\le0$ ($a\in[I+1,N]$).
The argument of $I$ is a point in $N$-dimensional integer space.

If a diagram contains a self-energy insertion into some internal line,
then 2 lines carry the same momentum $l$, and can be joined:
\begin{equation}
\raisebox{-5.8mm}{\begin{picture}(28,15.5)
\put(14,7.75){\makebox(0,0){\includegraphics{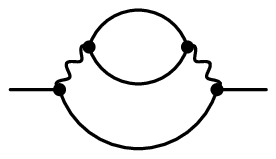}}}
\put(4.5,10){\makebox(0,0){$n_1$}}
\put(23.5,10){\makebox(0,0){$n_2$}}
\end{picture}} =
\raisebox{-5.8mm}{\begin{picture}(37,14.5)
\put(23,7.25){\makebox(0,0){\includegraphics{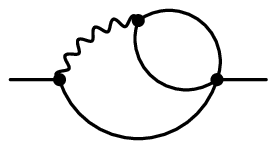}}}
\put(8.5,11){\makebox(0,0){$n_1+n_2$}}
\end{picture}} =
\raisebox{-5.8mm}{\begin{picture}(37,14.5)
\put(14,7.25){\makebox(0,0){\includegraphics{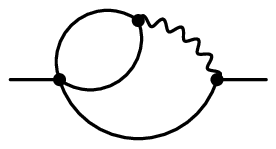}}}
\put(28,11){\makebox(0,0){$n_1+n_2$}}
\end{picture}}\,.
\label{Intro:Insert}
\end{equation}
These 3 different graphs correspond to 1 Feynman integral.

If the lines around a self-energy insertion have different mass
(e.\,g., $\gamma$ and $Z^0$), e.\,g.,
\begin{equation*}
\raisebox{-5.8mm}{\begin{picture}(28,15.5)
\put(14,7.75){\makebox(0,0){\includegraphics{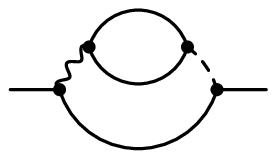}}}
\end{picture}}\,,
\end{equation*}
then their denominators are different but linear-dependent.
We shall see that it's easy to kill one of these lines
using partial-fraction decomposition.
It's best to do this before further calculations;
otherwise, expressing scalar products in the numerator via the denominators
becomes non-unique.
However, this is not possible if these lines are raised to non-integer powers
(see below).

If some subdiagram is connected to the rest of the diagram only at 1 vertex
(and has no external legs), it can be completely separated and considered
a vacuum diagram:
\begin{equation}
\raisebox{-11.3mm}{\begin{picture}(25,20)
\put(12.5,10){\makebox(0,0){\includegraphics{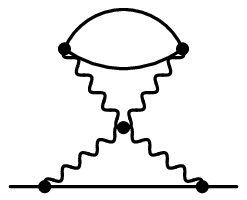}}}
\end{picture}} =
\raisebox{-11.3mm}{\begin{picture}(25,25)
\put(12.5,4){\makebox(0,0){\includegraphics{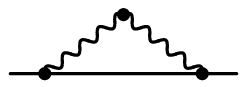}}}
\put(12.5,18){\makebox(0,0){\includegraphics{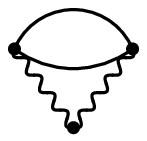}}}
\end{picture}}\,.
\label{Intro:c1}
\end{equation}
In particular,
\begin{equation}
\raisebox{-8.8mm}{\begin{picture}(25,20)
\put(12.5,10){\makebox(0,0){\includegraphics{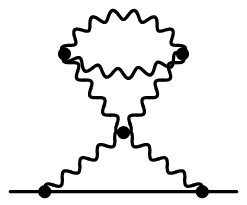}}}
\end{picture}} = 0\,.
\label{Intro:c10}
\end{equation}
If such a subdiagram has external legs, it still can be separated.
The momentum enters the rest of the diagram through a new external leg:
\begin{equation}
\raisebox{-15.55mm}{\begin{picture}(25,23.5)
\put(12.5,11.75){\makebox(0,0){\includegraphics{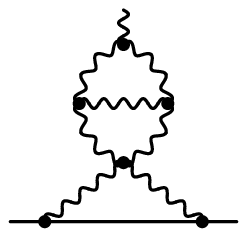}}}
\end{picture}} =
\raisebox{-15.55mm}{\begin{picture}(25,37.5)
\put(12.5,5.75){\makebox(0,0){\includegraphics{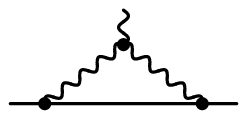}}}
\put(12.5,27){\makebox(0,0){\includegraphics{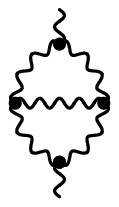}}}
\end{picture}}\,.
\label{Intro:c2}
\end{equation}
In particular,
\begin{equation}
\raisebox{-10.55mm}{\begin{picture}(25,23.5)
\put(12.5,11.75){\makebox(0,0){\includegraphics{touch5.eps}}}
\put(14,22){\makebox(0,0)[l]{$p^2=0$}}
\end{picture}} = 0\,.
\label{Intro:c20}
\end{equation}

Suppose some subdiagram is connected to the rest of the diagram only at 2 vertices,
and has no external legs (let's call this subdiagram $a$).
We can always choose the total momentum flowing through this subdiagram to be $k_1$.
Let the loop momenta $k_2$, \dots, $k_A$ live in the subdiagram $a$;
then the remaining loop momenta $k_{A+1}$, \dots, $k_L$
live in the remaining subdiagram $b$.
Suppose the total degree of the numerator in mixed scalar products $k_i\cdot q_j$
(with $i\le A$, $j>A$) is $n$.
Then the subdiagram $a$ is an integral in $k_2$, \dots, $k_A$ with $n$ tensor indices,
depending only on $k_1$.
If we detach this subdiagram and then attach it the other way round,
this means $k_1\to-k_1$.
Let's make the substitution $k_2\to-k_2$, \dots, $k_A\to-k_A$.
All the denominators in the subdiagram $a$ remain unchanged, and
\begin{equation}
\raisebox{-8.8mm}{\begin{picture}(38,20)
\put(19,10){\makebox(0,0){\includegraphics{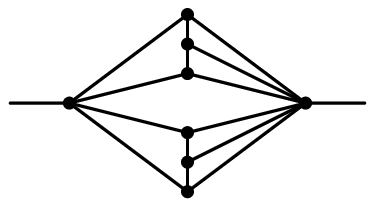}}}
\end{picture}} = (-1)^n
\raisebox{-8.8mm}{\begin{picture}(38,20)
\put(19,10){\makebox(0,0){\includegraphics{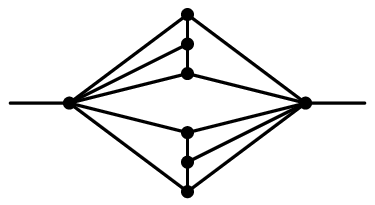}}}
\end{picture}}\,.
\label{Intro:c3}
\end{equation}
The second equality in~(\ref{Intro:Insert}) is a particular case of this property.

\begin{figure}[ht]
\begin{center}
\begin{picture}(76,76)
\put(37,37){\makebox(0,0){\includegraphics{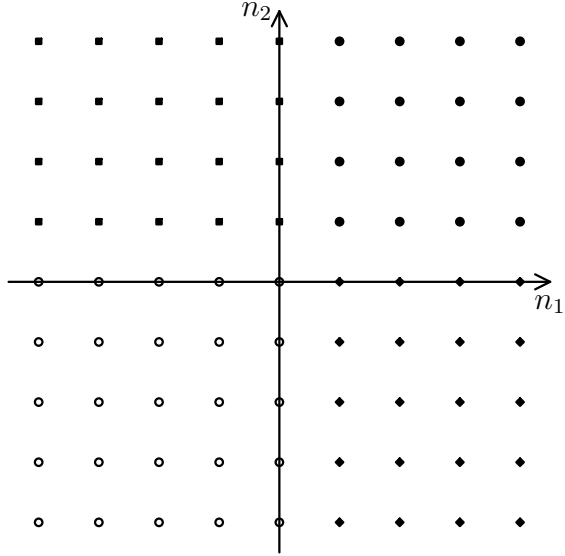}}}
\put(73,34){\makebox(0,0){$n_1$}}
\put(34,73){\makebox(0,0){$n_2$}}
\end{picture}
\end{center}
\caption{Sectors in the $N$-dimensional integer space}
\label{F:Sect}
\end{figure}

The $N$-dimensional integer space in which Feynman integrals~(\ref{Intro:I}) live
can be subdivided into sectors (Fig.~\ref{F:Sect}).
If $n_a>0$, $D_a$ is in the denominator;
if $n_a\le0$, it is in the numerator (the line is shrunk).
The set of sectors is partially ordered:
the $++$ sector is higher than $+-$ which is higher than $--$;
the $++$ sector is higher than $-+$ which is higher than $--$;
but the sectors $+-$ and $-+$ cannot be compared (Fig.~\ref{F:Sect}).
Each sector has a \textit{corner} --- the point with
$n_a=1$ (if $n_a>0$ in the sector) or $n_a=0$ (if $n_a\le0$ in the sector).
Generally speaking, there are $2^N$ sectors.
But for irreducible numerator $D_a$, we always have $n_a\le0$,
and opposite sectors don't exist.
Some sectors are trivial, i.\,e.\ $I=0$ in all points.
At least, the pure negative sector (where all $n_a\le0$) is trivial.
Often there are more trivial sectors, when shrinking some lines
produces a scale-free vacuum subdiagram~(\ref{Intro:c10})
(some sectors are trivial only at some specific kinematics,
e.\,g., (\ref{Intro:c20})).
In sectors just above trivial ones
(i.\,e.\ when the diagram vanishes after contracting any line)
a general expression for the integral (usually via $\Gamma$ functions)
can often be obtained.

Some sectors are transformed into each other by symmetries.
All the denominators of the original integral become
those of the symmetric integral,
possibly, after an appropriate integration-momenta substitution.
The numerators of the original integral (i.\,e., $D_a$  whose $n_a<0$),
after the symmetry transformtion and the loop-momenta substitution,
can be expressed as linear combinations of the numerators $D_a$ in the new sector.
Expanding the product of powers of such numerators,
we can express the original integral as a linear combination
of integrals in the new sector having different powers of the new numerators.
A sector may be transformed into itself by some symmetries.

Let's consider a self-energy insertion into a massless line~(\ref{Intro:line})
(the HQET propagator~(\ref{Intro:HQET}) is also ``massless'' in this sense,
because it contains no dimensional parameters except $l$).
If this insertion contains no massive lines, then, by dimensionality,
this insertion just shifts the power of the propagator:
\begin{equation}
\begin{split}
&\raisebox{-8.8mm}{\begin{picture}(40,20)
\put(20,10){\makebox(0,0){\includegraphics{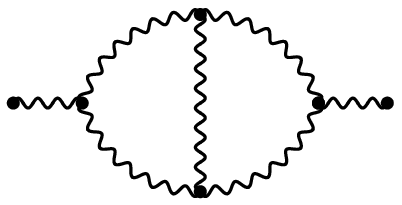}}}
\end{picture}} \Rightarrow
\raisebox{-0.3mm}{\begin{picture}(22,6)
\put(11,1.5){\makebox(0,0){\includegraphics{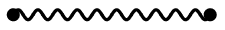}}}
\put(11,7){\makebox(0,0){$\displaystyle-L\frac{d}{2}+n$}}
\end{picture}}\,,\\
&\raisebox{-8.8mm}{\begin{picture}(40,20)
\put(20,10){\makebox(0,0){\includegraphics{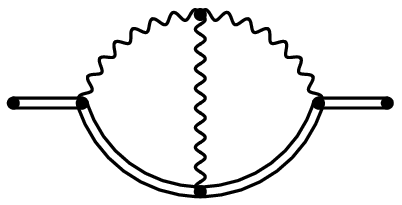}}}
\end{picture}} \Rightarrow
\raisebox{-0.3mm}{\begin{picture}(22,6)
\put(11,1.5){\makebox(0,0){\includegraphics{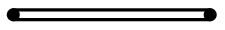}}}
\put(11,4.5){\makebox(0,0){$-Ld+n$}}
\end{picture}}\,,
\end{split}
\label{Intro:Ins0}
\end{equation}
where $L$ is the number of loops in the insertion,
and the integer $n$ is determined by the powers of the denominators in it.
A diagram with such insertion[s] can be considered as a lower-loop diagram~(\ref{Intro:I})
with non-integer ($d$-dependent) power[s] of some denominator[s].
Such a power cannot be compared to integers,
and hence there is no subdivision into sectors along the corresponding $n$ axis.
This factor is always in the denominator,
and can newer become a numerator.
I know no circumstances when a \emph{massive} propagator gets raised to a non-integer power%
\footnote{Except the situation when the $d$-dimensional space is separated into
an $n$-dimensional subspace and a transverse $(d-n)$-dimensional one;
integrating a massive propagator in transverse momentum components
produces a massive propagator in the $n$-dimensional subspace
raised to a non-integer power.}.

\section{Integration by parts}
\label{S:IBP}

The Feynman integral~(\ref{Intro:I}) does not change if we do a substitution
\begin{equation}
k_i \to M_{ij} q_j = A_{ij} k_j + B_{ij} p_j
\label{IBP:M}
\end{equation}
of its integration momenta, where
\begin{equation*}
M = \left(
\begin{array}{cccccc}
A_{11}  & \cdots & A_{1L}  & B_{11} & \cdots & B_{1E} \\
\vdots & \ddots & \vdots & \vdots & \ddots & \vdots \\
A_{L1}  & \cdots & A_{LL}  & B_{L1} & \cdots & B_{LE}
\end{array}
\right)
\end{equation*}
is an $L\times M$ matrix, provided that the substitution is invertible:
\begin{equation}
\det A \ne 0\,.
\label{IBP:det}
\end{equation}
Such substitutions form a Lie symmetry group of the Feynman integral.

Let's consider an infinitesimal transformation $k_i\to k_i+\alpha q_j$.
The integrand transforms as
\begin{equation}
f \to f + \alpha q_j \cdot \partial_i f\,.
\label{IBP:f}
\end{equation}
If $j=i$, also the integration measure changes:
\begin{equation}
d^d k_i \to (1 + \alpha d) d^d k_i\,.
\label{IBP:dk}
\end{equation}
These  infinitesimal transformations form the Lie algebra~\cite{L:08}
\begin{equation}
\begin{split}
&\int d^d k_1 \cdots d^d k_L\,O_{ij} f = 0\,,\\
&O_{ij} = \partial_i \cdot q_j\qquad
(i\le L,\;j\ge i)\,,\qquad
\partial_i = \frac{\partial}{\partial q_i}\,.
\end{split}
\label{IBP:Lie}
\end{equation}
The generators obey the commutation relation
\begin{equation}
[O_{ij},O_{i'j'}] = \delta_{ij'} O_{i'j} - \delta_{i'j} O_{ij'}
\label{IBP:comm}
\end{equation}
(structure constants).

Let's find an explicit form of the operator
\begin{equation}
\begin{split}
&O_{ij} = d \delta_{ij} + q_j \cdot \partial_i
= d \delta_{ij} + \sum_{m=1}^M (1+\delta_{mi}) s_{mj} \frac{\partial}{\partial s_{mi}}\,,\\
&\frac{\partial}{\partial s_{mi}} = \sum_{a=1}^N A_a^{mi} \frac{\partial}{\partial D_a}
\end{split}
\label{IBP:Oij}
\end{equation}
(the extra factor 2 at $m=i$ comes from $\partial_i s_{ii} = 2 k_i$).
We assume that whenever $s_{ij}$ with $i>j$ appears in an equation,
it is immediately replaced by $s_{ji}$;
the same holds for $\partial/\partial s_{ij}$, $A^a_{ij}$, $A_a^{ij}$.
If $j\le L$ ($q_j=k_j$), then from~(\ref{Intro:sD}) we have
\begin{equation}
O_{ij} = d \delta_{ij}
+ \sum_{a=1}^N \sum_{b=1}^N \sum_{m=1}^M A_a^{mi} A^b_{mj} (1+\delta_{mi})
(D_b - m_b^2) \frac{\partial}{\partial D_a}\,;
\label{IBP:IBPk}
\end{equation}
if $j>L$ ($q_j=p_{j-L}$), then $s_{mj}$ with $m,j>L$ are external kinematic quantities:
\begin{equation}
O_{ij} = \sum_{a=1}^N \left[
\sum_{m=1}^L \sum_{b=1}^N A_a^{mi} A^b_{mj} (1+\delta_{mi}) (D_b - m_b^2)
+ \sum_{m=L+1}^M A_a^{mi} s_{mj} \right] \frac{\partial}{\partial D_a}\,.
\label{IBP:IBPp}
\end{equation}
The operator $\partial/\partial D_a$ acting on the integrand $f$~(\ref{Intro:I})
raises the power $n_a$ by 1 (and multiplied the integrand by $n_a$);
$D_b$ lowers $n_b$ by 1.

Let's introduce operators which act on functions of $N$ integer variables
and produce new functions:
\begin{equation}
\begin{split}
(\mathbf{n}_a F) (n_1,\ldots,n_a,\ldots,n_N) &{}= n_a F(n_1,\ldots,n_a\ldots,n_N)\,,\\
(\mathbf{a}^+ F) (n_1,\ldots,n_a,\ldots,n_N) &{}= F(n_1,\ldots,n_a+1,\ldots,n_N)\,,\\
(\mathbf{a}^- F) (n_1,\ldots,n_a,\ldots,n_N) &{}= F(n_1,\ldots,n_a-1,\ldots,n_N)\,.
\end{split}
\label{IBP:Oper}
\end{equation}
The shift operators are inverse to each other:
\begin{equation}
\mathbf{a}^+ \mathbf{a}^- = \mathbf{a}^- \mathbf{a}^+ = 1\,.
\label{IBP:inv}
\end{equation}
They don't commute with the number operators $\mathbf{n}_a$:
\begin{equation}
[\mathbf{a}^\pm,\mathbf{n}_b] = \pm \delta_{ab} \mathbf{a}^\pm\,.
\label{IBP:comm2}
\end{equation}

Some authors prefer to use the operators $\hat{\mathbf{a}}^+ = \mathbf{n}_a \mathbf{a}^+$,
\begin{equation}
(\hat{\mathbf{a}}^+ F) (n_1,\ldots,n_a,\ldots,n_N) = n_a F(n_1,\ldots,n_a+1,\ldots,n_N)
\label{IBP:hat}
\end{equation}
instead of $\mathbf{a}^+$, because
\begin{equation}
\frac{\partial}{\partial D_a} \Rightarrow - \hat{\mathbf{a}}^+\,,\qquad
D_b \Rightarrow \mathbf{b}^-
\label{IBP:subs}
\end{equation}
are the only combinations appearing in~(\ref{IBP:IBPk}), (\ref{IBP:IBPp}).
These operators obey the commutation relation
\begin{equation}
[\hat{\mathbf{a}}^+,\mathbf{b}^-] = \delta_{ab}\,.
\label{IBP:comm3}
\end{equation}
Then the operators $\mathbf{n}_a = \hat{\mathbf{a}}^+ \mathbf{a}^-$
are not independent; it is sufficient to use
$\hat{\mathbf{a}}^+$ and $\mathbf{a}^-$.

Now the identities~(\ref{IBP:Lie}) can be rewritten as the IBP recurrence relations
\begin{equation}
\mathbf{O}_{ij}(\hat{\mathbf{a}}^+,\mathbf{a}^-) I(n_1,\ldots,n_N) =
\frac{1}{(i\pi^{d/2})^L} \int  d^d k_1 \cdots d^d k_L\,O_{ij} f = 0\,,
\label{IBP:IBP}
\end{equation}
where the operators $\mathbf{O}_{ij}$ are obtained from~(\ref{IBP:IBPk}), (\ref{IBP:IBPp})
using the substitutions~(\ref{IBP:subs}):
\begin{align}
&\mathbf{O}_{ij} = d \delta_{ij}
- \sum_{a=1}^N \sum_{b=1}^N \sum_{m=1}^M A_a^{mi} A^b_{mj} (1+\delta_{mi})
\hat{\mathbf{a}}^+ \left(\mathbf{b}^- - m_b^2\right)\qquad
(i\le L)\,,
\label{IBP:k}\\
&\mathbf{O}_{ij} = \sum_{a=1}^N \left[ \sum_{b=1}^N \sum_{m=1}^L A_a^{mi} A^b_{mj} (1+\delta_{mi})
\hat{\mathbf{a}}^+ \left(\mathbf{b}^- - m_b^2\right)
- \sum_{m=L+1}^M A_a^{mi} s_{mj} \hat{\mathbf{a}}^+ \right]\qquad
(i>L)
\label{IBP:p}
\end{align}
(note the order of $\hat{\mathbf{a}}^+$ and $\mathbf{b}^-$).
The operators $\mathbf{O}_{ij}(\hat{\mathbf{a}}^+,\mathbf{a}^-)$
obey the commutation relations~(\ref{IBP:comm}) by virtue of~(\ref{IBP:comm3}).

\begin{figure}[ht]
\begin{center}
\begin{picture}(114,35)
\put(16,16){\makebox(0,0){\includegraphics{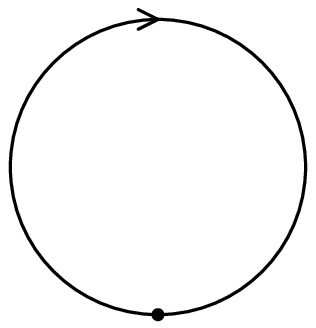}}}
\put(16,33.5){\makebox(0,0)[b]{$k$}}
\put(16,29.5){\makebox(0,0)[t]{$n$}}
\put(77,16){\makebox(0,0){\includegraphics{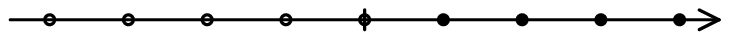}}}
\put(113,13){\makebox(0,0){$n$}}
\end{picture}
\end{center}
\caption{The 1-loop vacuum diagram and its sectors}
\label{F:v1}
\end{figure}

The simplest example is the 1-loop vacuum diagram (Fig.~\ref{F:v1})
\begin{equation}
\frac{1}{i\pi^{d/2}} \int \frac{d^d k}{D^n} = V(n) m^{d-2n}\,,\qquad
D = m^2 - k^2 - i0\,.
\label{IBP:V1}
\end{equation}
We may set $m=1$; the power of $m$ can be reconstructed by dimensionality.
The sector $n\le0$ is trivial (Fig.~\ref{F:v1}).
The IBP relation
\begin{equation}
(d - 2 n + 2 n \mathbf{1}^+ ) V(n) = 0
\label{IBP:V1rel}
\end{equation}
relate 2 neighbouring values of $n$
(except the $n=0$ relation where $V(1)$ does not appear,
and we obtain the known fact $V(0)=0$).
In the positive sector, we can express $V(n)$ via $V(n-1)$,
then $V(n-2)$, and so on, until we reach $V(1)$.
The explicit solution of~(\ref{IBP:V1rel}) is
\begin{equation}
V(n) = \frac{1}{\Gamma(n)}
\frac{\Gamma\left(n-\frac{d}{2}\right)}{\Gamma\left(1-\frac{d}{2}\right)}
V(1)\,.
\label{IBP:V1sol}
\end{equation}

\begin{figure}[ht]
\begin{center}
\begin{picture}(108,76)
\put(11,37){\makebox(0,0){\includegraphics{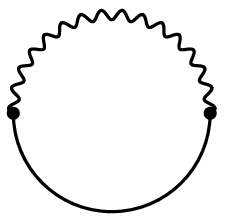}}}
\put(11,24.5){\makebox(0,0){1}}
\put(11,49.5){\makebox(0,0){2}}
\put(69,37){\makebox(0,0){\includegraphics{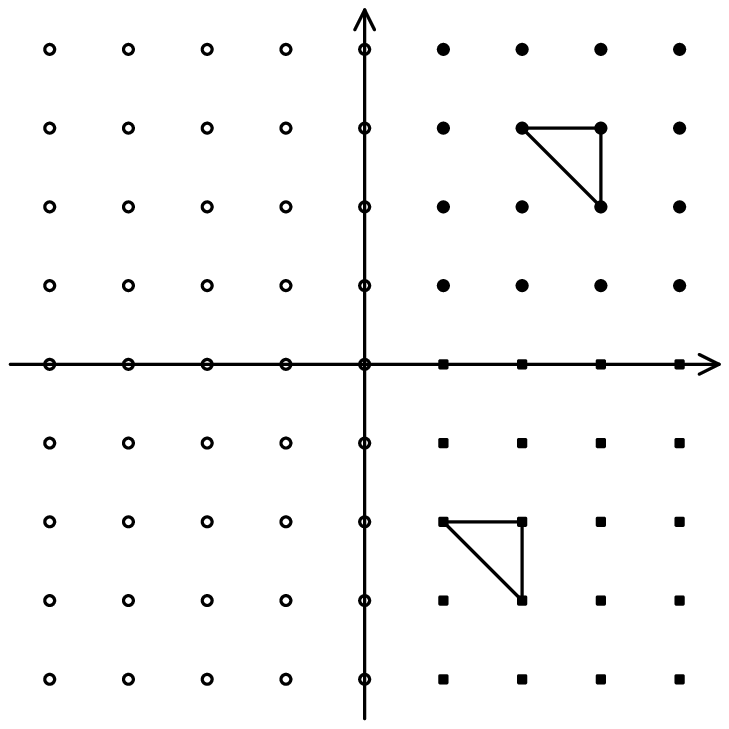}}}
\put(105,34){\makebox(0,0){$n_1$}}
\put(66,73){\makebox(0,0){$n_2$}}
\end{picture}
\end{center}
\caption{The 1-loop vacuum diagram with masses $m$, 0 and its sectors}
\label{F:m0}
\end{figure}

The 1-loop vacuum diagram with masses $m$ and 0 (Fig.~\ref{F:m0})
contains linearly dependent denominators
\begin{equation}
D_1 = 1 - k^2\,,\qquad
D_2 = - k^2\,,\qquad
D_1 - D_2 = 1
\label{IBP:m0den}
\end{equation}
(we set $m=1$).
In the positive sector, we solve the recurrence relation
\begin{equation}
(1 - \mathbf{1}^- + \mathbf{2}^-) I = 0
\label{IBP:m0rel}
\end{equation}
for $I$; each reduction step decreases $n_1+n_2$ by 1,
and eventually we get $n_1=0$ (where $I=0$) or $n_2=0$.
In the sector $n_2<0$, we solve for $\mathbf{2}^- I$;
each step decreases $|n_2|$ by 1, and we end up
at $n_2=0$ or $n_1=0$ (Fig.~\ref{F:m0}).
This reduction is equivalent to partial fraction decomposition.
So, this problem reduces to the previous one (Fig.~\ref{F:v1})
(unless $n_2$ is not integer).

\begin{figure}[ht]
\begin{center}
\begin{picture}(98,20)
\put(17,11.5){\makebox(0,0){\includegraphics{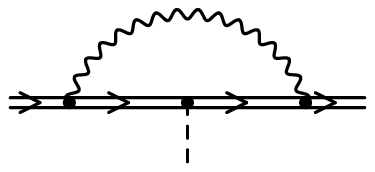}}}
\put(77,10){\makebox(0,0){\includegraphics{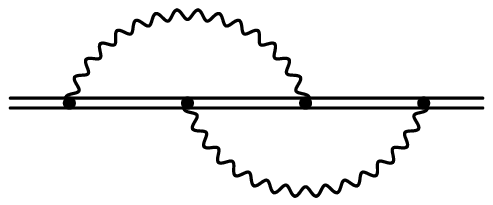}}}
\end{picture}
\end{center}
\caption{HQET diagrams with linearly dependent denominators}
\label{F:HQETpf}
\end{figure}

Linear dependent denominators often appear in HQET diagrams~\cite{BG:91}.
For example, the vertex diagram in Fig.~\ref{F:HQETpf} has
\begin{align*}
&D_1 = - 2 (k+p_1) \cdot v = - 2 (k\cdot v + \omega_1)\,,\qquad
D_2 = - 2 (k+p_2) \cdot v = - 2 (k\cdot v + \omega_2)\,,\\
&D_1 - D_2 + 2 (\omega_1 - \omega_2) = 0\,.
\end{align*}
The recurrence relation
\begin{equation}
\left[ 2(\omega_1-\omega_2) + \mathbf{1}^- - \mathbf{2}^- \right] I = 0
\label{IBP:HQETpf}
\end{equation}
allows one to kill one of the heavy lines, similarly to Fig.~\ref{F:m0}.
If an $L$-loop diagram contains more than $L$ HQET lines (with the same $v$),
then its HQET denominators are linearly dependent:
there are only $L$ scalar products $k_i\cdot v$
(see, e.\,g., Fig.~\ref{F:HQETpf}).
Some of these HQET lines can be easily killed by partial fraction decomposition,
unless their powers are non-integer.

\begin{figure}[ht]
\begin{center}
\begin{picture}(124,76)
\put(19,37){\makebox(0,0){\includegraphics{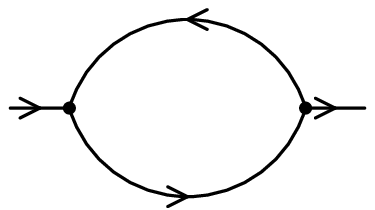}}}
\put(19,25.5){\makebox(0,0){$k+p$}}
\put(19,48.5){\makebox(0,0){$k$}}
\put(4,34.5){\makebox(0,0){$p$}}
\put(34,34.5){\makebox(0,0){$p$}}
\put(85,37){\makebox(0,0){\includegraphics{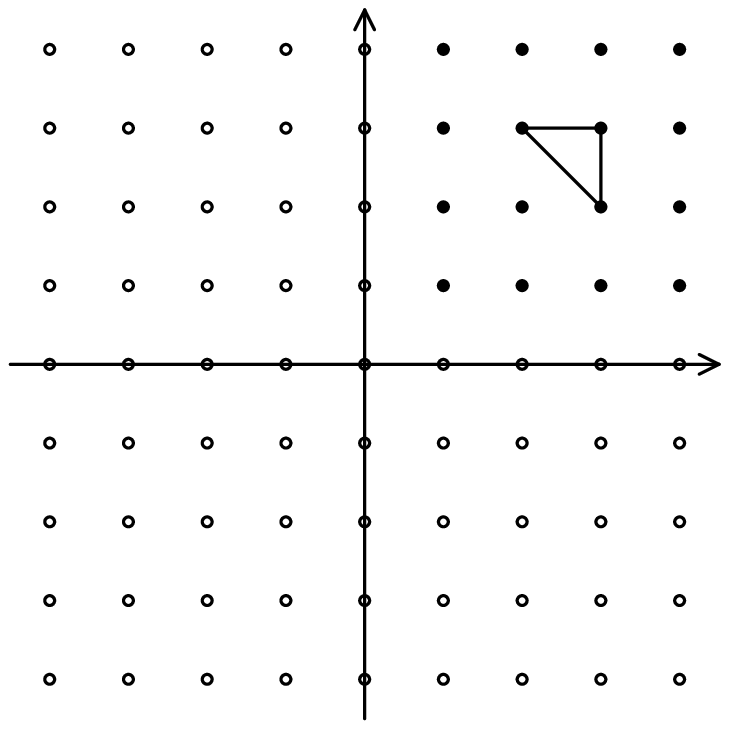}}}
\put(121,34){\makebox(0,0){$n_1$}}
\put(82,73){\makebox(0,0){$n_2$}}
\end{picture}
\end{center}
\caption{The 1-loop massless self-energy diagram and its sectors}
\label{F:p1}
\end{figure}

The 1-loop massless self-energy diagram (Fig.~\ref{F:p1}) has
\begin{equation*}
D_1 = - (k+p)^2\,,\qquad
D_2 = -k^2\,.
\end{equation*}
It has only one non-trivial sector,
and is symmetric with respect to $1\leftrightarrow2$.
We may put $p^2=-1$ (its power can be restored by dimensionality).
The $\partial\cdot k$ IBP relation is
\begin{equation}
\left[d - n_1 - 2 n_2 + n_1 \mathbf{1}^+ (1 - \mathbf{2}^-)\right] G = 0\,.
\label{IBP:p1}
\end{equation}
If $n_1>1$, we can use~(\ref{IBP:p1}) to lower $n_1+n_2$ by one;
if $n_2>1$, the mirror-symmetric $\partial\cdot(k-p)$ relation can be used instead.
All integrals reduce to 1 master integral $G(1,1)$%
\footnote{Reduction for the 1-loop massless triangle diagram
has been considered in~\cite{D:92}.}.

In general, when solving the reduction problem in some sector,
integrals from lower sectors (where some lines are contracted) are considered trivial,
i.\,e., reduction problems in those sectors are assumed to be solved.
In other words, the reduction problem is recursive:
to solve it for the fully positive sector,
one has to solve it for all the sectors just below it,
and so on, until we reach trivial sectors (where $I=0$).

Automatic identification of trivial sectors is a useful feature
for programs of IBP reduction.
If there exists a substitution $M(\alpha)$~(\ref{IBP:M})
such than the integral is multiplied by $\alpha^n$
(where $n\ne0$ is $d$-dependent), then the integral obviously vanishes
(contains a scale-free subdiagram).
The same holds if any numerator is inserted into this integral.
Therefore, if the integral at the corner of the sector vanishes,
the whole sector is trivial.
This gives the criterion~\cite{L:08}:
if solving all IBP relations at the corner point results in $I=0$,
the sector is trivial.
This criterion finds all cases like~(\ref{Intro:c10}),
but not sectors which are trivial only in specific kinematics,
like~(\ref{Intro:c20}).

When considering some specific sector,
we should express more complicated integrals via simpler ones.
To do this systematically, we need to accept some convention
which integrals are more complicated and which are more simple.
This means introducing a total order of integer points in the sector.
It is natural to require this order to be translation invariant:
if $\vec{n}_1 < \vec{n}_2$ then $\vec{n}_1+\vec{m} < \vec{n}_2+\vec{m}$.
The choice of such an admissible order is, of course, not unique.
We want the reduction process to move integrals
closer to the corner of the sector.
Therefore, it is reasonable to compare the sums $n = \sum_a |n_a|$
first: if the first integral has a higher $n$ than the second one,
the first integral is more complicated.
After that, we should decide how to order integrals with equal values of $n$.
This can be done in different ways (e.\,g., lexicographically).

Suppose we have fixed some admissible order in the sector.
Let's peek some IBP relation and solve it for the most complicated integral.
This solution can be used for reduction of all integrals in the sector.
They either move into lower sectors
(when some $n_a=1$ and $\mathbf{a}^-$ is applied),
or end up in a hyperplane of dimensionality $N-1$ in our sector.
Indeed, the most complicated integral in our IBP relation is
$n_a \mathbf{a}^+ I$ for some $a$;
we cannot reduce $n_a$ from 1 to 0 using this relation.
So, we are left with the hyperplane $n_a=1$.

Next we update the remaining IBP relations
to contain only integrals on this hyperplane
(and trivial integrals from lower sectors:
$\mathbf{a}^-$ acting on an integral with $n_a=1$ makes it trivial).
Whenever an IBP relation contains $n_a=2$,
we lower it back to $n_a=1$ using the first selected relation.
Now the problem is much simpler:
the dimensionality of space is $N-1$ instead of $N$.
Let's select one of those $(N-1)$-dimensional recurrence relations,
and solve it for the most complicated integral.
Then we can use this solution to further reduce all integrals
on the hyperplane.
Here is a catch, however:
the coefficient of this most complicated integral may vanish
on some complicated subset of points,
and the reduction fails on this subset
(this can never happen during the first stage:
there this subset is always the hyperplane $n_a=1$).
This subset is still much smaller than the whole hyperplane:
it consists of some lower-dimensional part[s].
It is better to choose a relation for which this subset is simple
(e.\,g., a coordinate hyperplane).

Next we treat these $(N-2$)-dimensional subsets in a similar way, and so on.
This is a sketch of what people usually do
when constructing a reduction algorithm by hand.
This strategy is implemented in a \texttt{\textit{Mathematica}} program
by R.\,N.~Lee based on~\cite{L:08}.
This program is not guaranteed to construct a reduction algorithm
for a given topology.
However, it works successfully (and efficiently)
for a large number of highly non-trivial examples.
It is not publicly available.

In some rare cases, approaches based on sectors can fail to detect a relation
between 2 integrals in some sector, and declare both to be masters,
when in fact they are dependent, but this dependency can only be obtained
via a higher sector.

\section{Homogeneity and Lorentz-invariance relations}
\label{S:Add}

Scalar Feynman integrals obey some relations
which can be derived independently of IBP
but are not really independent ---
they appear to be linear combinations of the IBP relations.

By dimensionality, an $L$-loop integral $I$ is a homogeneous function
of external momenta $p_i$ and masses $m_i$ of degree $Ld - 2\sum n_i$
(we assume that all the denominators are quadratic).
Therefore,
\begin{equation}
\left( \sum_i p_i \cdot \frac{\partial}{\partial p_i}
+ \sum_i m_i \frac{\partial}{\partial m_i} \right) I
= \left( L d - 2 \sum_i n_i \right) I\,.
\label{Add:Hom0}
\end{equation}
On the other hand, the derivatives in the left-hand side
can be calculated explicitly.
Equation these two expressions,
we obtain a homogeneity relation~\cite{CT:81}.

Let's move all the terms in~(\ref{Add:Hom0}) to the left-hand side,
and apply this operator to the integrand $f$ instead of the integral $I$.
Adding and subtracting the sum over the loop momenta $k_i$ we get
\begin{align*}
&\left( \sum_i  p_i \cdot \frac{\partial}{\partial p_i}
+ \sum_i m_i \frac{\partial}{\partial m_i}
- L d + 2 \sum_i n_i \right) f\\
&{} = \left( \sum_i  q_i \cdot \frac{\partial}{\partial q_i}
+ \sum_i m_i \frac{\partial}{\partial m_i}
- \sum_i  k_i \cdot \frac{\partial}{\partial k_i}
- L d + 2 \sum_i n_i \right) f\,.
\end{align*}
The integrand $f$ is a homogeneous function of the momenta $q_i$
(both loop and external) and the masses $m_i$ of degree $-2\sum n_i$,
and this expression simplifies to
\begin{equation}
\left( - \sum_i  k_i \cdot \frac{\partial}{\partial k_i} - L d \right) f =
- \left( \sum_i \frac{\partial}{\partial k_i} \cdot k_i \right) f\,.
\label{Add:Hom1}
\end{equation}
Thus the homogeneity relation is a linear combination of the IBP relations
$\sum \partial_i\cdot k_i$.

A scalar integral $I$ does not change if we rotate the external momenta $p_i$.
In other words, a Lorentz-transformation generator applied to $I$ gives 0.
If there are at least 2 external momenta ($E\ge2$),
we can contract this tensor equation with $p_i^\mu p_j^\nu$ ($i\ne j$)
to obtain a scalar relation
\begin{equation}
2 p_i^\mu p_j^\nu \left( \sum_n p_n^{[\mu} \frac{\partial}{\partial p_n^{\nu]}} \right) I = 0
\label{Add:LI0}
\end{equation}
(square brackets mean antisymmetrization).
On the other hand, the derivatives can be calculated explicitly.
This gives Lorentz-invariance relations~\cite{GR:99}.

Let's apply this operator to the integrand $f$.
Adding and subtracting the sum over the loop momenta $k_i$ we get
\begin{equation*}
2 p_i^\mu p_j^\nu \left( \sum_n p_n^{[\mu} \frac{\partial}{\partial p_n^{\nu]}} \right) f
= 2 p_i^\mu p_j^\nu \left( \sum_n q_n^{[\mu} \frac{\partial}{\partial q_n^{\nu]}}
- \sum_n k_n^{[\mu} \frac{\partial}{\partial k_n^{\nu]}} \right) f\,.
\end{equation*}
The integrand $f$ is a scalar function of the momenta $q_i$,
and hence the first operator gives 0 when acting on $f$.
Writing the antisymmetrization explicitly and commuting derivatives to the left,
we obtain
\begin{equation}
\begin{split}
- 2 p_i^\mu p_j^\nu \left( \sum_n  k_n^{[\mu} \frac{\partial}{\partial k_n^{\nu]}} \right) f
&{} = \sum_n \left( p_j\cdot k_n\,p_i\cdot\frac{\partial}{\partial k_n}
- p_i\cdot k_n\,p_j\cdot\frac{\partial}{\partial k_n} \right) f\\
&{} = \sum_n \frac{\partial}{\partial k_n} \cdot
\left( p_i\,p_j\cdot k_n - p_j\,p_i\cdot k_n \right) f
\end{split}
\label{Add:LI1}
\end{equation}
(the extra terms from commutation cancel).
Thus Lorentz-invariance relations are linear combinations of IBP relations~\cite{L:08}.

\section{Massless self-energy diagrams}
\label{S:p2}

\begin{figure}[ht]
\begin{center}
\begin{picture}(64,32)
\put(32,18){\makebox(0,0){\includegraphics{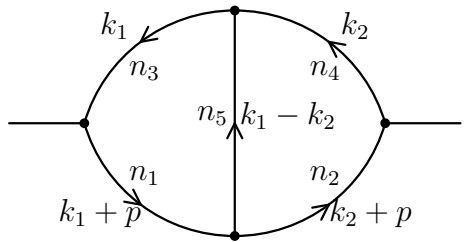}}}
\put(16,30){\makebox(0,0){$k_1$}}
\put(48,30){\makebox(0,0){$k_2$}}
\put(14,5){\makebox(0,0){$k_1+p$}}
\put(50,5){\makebox(0,0){$k_2+p$}}
\put(39,18){\makebox(0,0){$k_1-k_2$}}
\put(20,10){\makebox(0,0){$n_1$}}
\put(44,10){\makebox(0,0){$n_2$}}
\put(20,24){\makebox(0,0){$n_3$}}
\put(44,24){\makebox(0,0){$n_4$}}
\put(29,18){\makebox(0,0){$n_5$}}
\end{picture}
\end{center}
\caption{Two-loop massless self-energy diagram (all lines are massless)}
\label{F:Q2}
\end{figure}

Let's consider the 2-loop integral~\cite{CT:81}%
\footnote{The IBP and homogeneity relations for this integral
had been derived in~\cite{VPK:81} slightly earlier than in~\cite{CT:81};
however, the reduction algorithm had not been formulated.}
(Fig.~\ref{F:Q2})
\begin{equation}
\begin{split}
&\frac{1}{(i\pi^{d/2})^2} \int d^d k_1 d^d k_2\,f(k_1,k_2,p) =
G(n_1,\ldots,n_5) (-p^2)^{d/2-n_1-\cdots-n_5}\,,\\
&f(k_1,k_2,p) = \frac{1}{D_1^{n_1}\cdots D_5^{n_5}}\,,\\
&D_1=-(k_1+p)^2\,,\qquad
D_2=-(k_2+p)^2\,,\qquad
D_3=-k_1^2\,,\qquad
D_4=-k_2^2\,,\\
&D_5=-(k_1-k_2)^2\,.
\end{split}
\label{p2:G}
\end{equation}
It is symmetric with respect to the interchanges
$(1\leftrightarrow2,3\leftrightarrow4)$ and $(1\leftrightarrow3,2\leftrightarrow4)$.
It vanishes when indices of two adjacent lines are non-positive integers,
because then it contains a no-scale subdiagram.

When one of the indices is zero, the problem becomes trivial.
If $n_5=0$, it is the product of two one-loop diagrams:
\begin{equation}
\raisebox{-10.3mm}{\begin{picture}(52,23)
\put(26,11.5){\makebox(0,0){\includegraphics{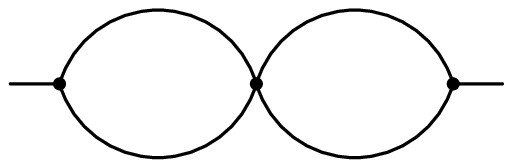}}}
\put(16,1){\makebox(0,0)[b]{$n_1$}}
\put(36,1){\makebox(0,0)[b]{$n_2$}}
\put(16,22){\makebox(0,0)[t]{$n_3$}}
\put(36,22){\makebox(0,0)[t]{$n_4$}}
\end{picture}}\,.
\label{p2:n5}
\end{equation}
If $n_1=0$, we first calculate the inner loop
(it shifts the power of the upper propagator in the outer loop),
and then the outer one:
\begin{equation}
\raisebox{-10.55mm}{\begin{picture}(32,23.5)
\put(16,11.5){\makebox(0,0){\includegraphics{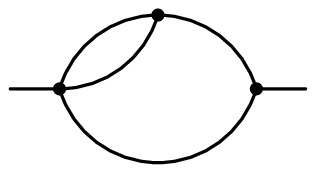}}}
\put(16,1){\makebox(0,0)[b]{$n_2$}}
\put(8,19){\makebox(0,0){$n_3$}}
\put(24,19){\makebox(0,0){$n_4$}}
\put(15,12.5){\makebox(0,0){$n_5$}}
\end{picture}} =
\raisebox{-10.55mm}{\begin{picture}(32,23.5)
\put(16,11.5){\makebox(0,0){\includegraphics{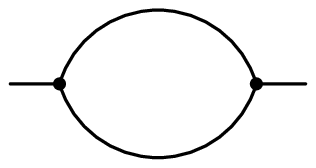}}}
\put(16,1){\makebox(0,0)[b]{$n_5$}}
\put(16,22.5){\makebox(0,0)[t]{$\vphantom{d}n_3$}}
\end{picture}} \times
\raisebox{-10.55mm}{\begin{picture}(32,23.5)
\put(16,11.5){\makebox(0,0){\includegraphics{q1.eps}}}
\put(16,1){\makebox(0,0)[b]{$n_2$}}
\put(16,22.5){\makebox(0,0)[t]{$n_4+n_3+n_5-d/2$}}
\end{picture}}\,.
\label{p2:n1}
\end{equation}
The cases $n_2=0$, $n_3=0$, $n_4=0$ are symmetric.
If all $n_i$ are integer,
then all integrals~(\ref{p2:n5}) are proportional to $G_1^2$,
and all integrals~(\ref{p2:n1}) to $G_2$:
\begin{equation}
\raisebox{-7.6mm}{\includegraphics{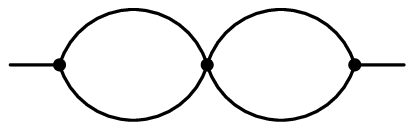}} = G_1^2\,,\qquad
\raisebox{-7.6mm}{\includegraphics{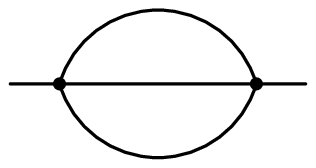}} = G_2\,,
\label{p2:masters}
\end{equation}
where
\begin{equation}
G_n =
\raisebox{-11.8mm}{\begin{picture}(44,26)
\put(22,13){\makebox(0,0){\includegraphics{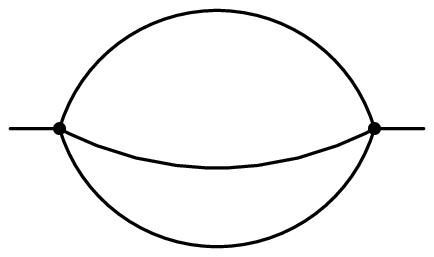}}}
\put(22,17){\makebox(0,0){$\cdots$}}
\end{picture}} =
\frac{\Gamma\left(n+1-n\frac{d}{2}\right)
\Gamma^{n+1}\left(\frac{d}{2}-1\right)}%
{\Gamma\left((n+1)\left(\frac{d}{2}-1\right)\right)}
\label{p2:Gn}
\end{equation}
is the $n$-loop massless sunset integral.

\begin{figure}[ht]
\begin{center}
\begin{picture}(150,50)
\put(25,25){\makebox(0,0){\includegraphics[width=50mm]{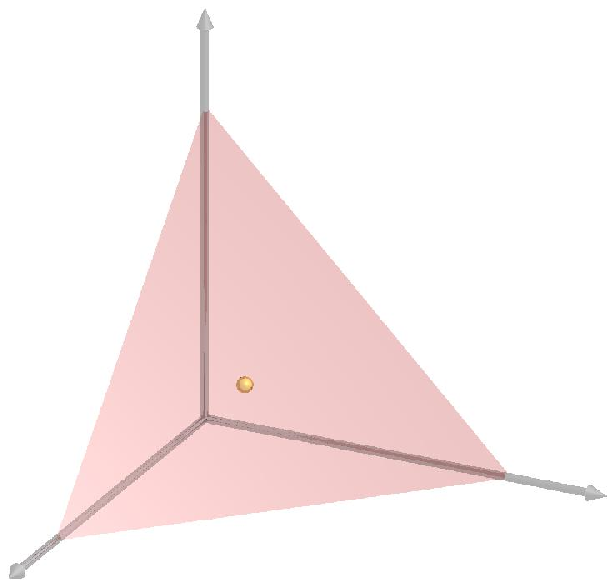}}}
\put(75,25){\makebox(0,0){\includegraphics[width=50mm]{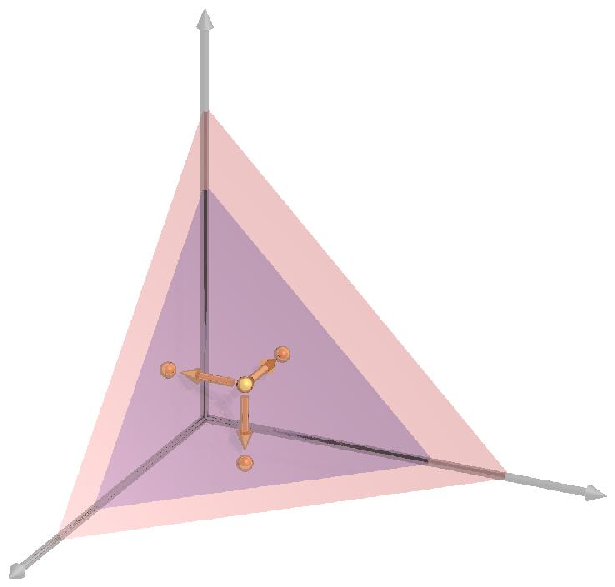}}}
\put(125,25){\makebox(0,0){\includegraphics[width=50mm]{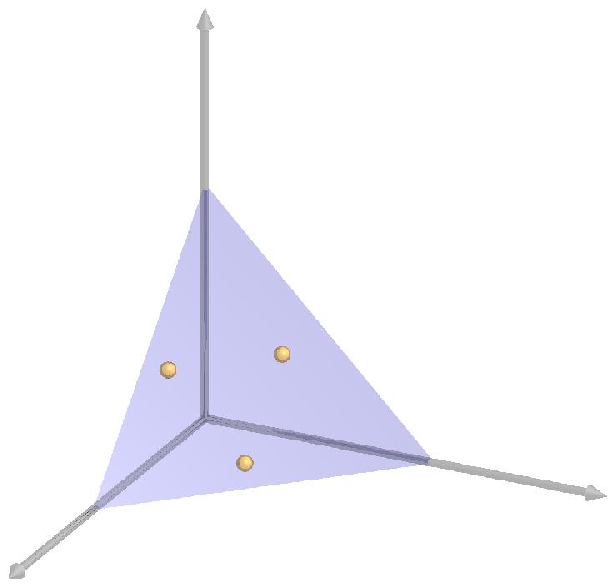}}}
\end{picture}
\end{center}
\caption{One step of IBP reduction: projection onto the $n_1$, $n_3$, $n_5$ subspace}
\label{F:IBP}
\end{figure}

Now we shall consider the positive sector (all $n_i>0$).
The $\partial_2\cdot k_2$ IBP relation is
\begin{equation}
[ d-n_2-n_5-2n_4 + n_2 \mathbf{2}^+ (1 - \mathbf{4}^-)
+ n_5 \mathbf{5}^+ (\mathbf{3}^- - \mathbf{4}^-) ] G = 0\,;
\label{p2:tri1}
\end{equation}
the $\partial_1\cdot k_1$ relation is mirror-symmetric.
The $\partial_2\cdot(k_2-k_1)$ IBP relation is
\begin{equation}
[d-n_2-n_4-2n_5 + n_2 \mathbf{2}^+ (\mathbf{1}^- - \mathbf{5}^-)
+ n_4 \mathbf{4}^+ (\mathbf{3}^- - \mathbf{5}^-)] G = 0\,.
\label{p2:tri2}
\end{equation}
Let's express $G(n_1,\ldots,n_5)$ (with unshifted indices) from this last expression.
Each application of this relation reduces $n_1+n_3+n_5$ by 1 (Fig.~\ref{F:IBP}).
Therefore, sooner or later one of the indices $n_1$, $n_3$, $n_5$ will vanish,
and we'll get a trivial case~(\ref{p2:n5}), (\ref{p2:n1}), or symmetric to it.
This means that all integrals in this sector
reduce to 2 master integrals~(\ref{p2:masters}).
Analyses of the remaining non-zero sectors is simple but somewhat lengthy;
all integrals in these sectors reduce to~(\ref{p2:masters}) too.

Applying $p\cdot(\partial/\partial p)$ to~(\ref{p2:G})
we get the homogeneity relation
\begin{equation}
[ 2(d-n_3-n_4-n_5)-n_1-n_2 + n_1 \mathbf{1}^+ (1 - \mathbf{3}^-)
+ n_2 \mathbf{2}^+ (1 - \mathbf{4}^-) ] G = 0\,.
\label{p2:hom}
\end{equation}
It is nothing but the sum of the $\partial_2\cdot k_2$ relation~(\ref{p2:tri1})
and its mirror-symmetric $\partial_1\cdot k_1$ relation.

Another interesting relation is obtained by inserting $(k_1+p)^\mu$
into the integrand of~(\ref{p2:G}) and taking derivative $\partial/\partial p^\mu$
of the integral.
On the one hand, the vector integral must be proportional to $p^\mu$,
and we can make the substitution
\begin{equation*}
k_1+p \to \frac{(k_1+p)\cdot p}{p^2} p =
\left(1 + \frac{D_1-D_3}{-p^2}\right) \frac{p}{2}
\end{equation*}
in the integrand.
Taking $\partial/\partial p^\mu$ of this vector integral produces~(\ref{p2:G}) with
\begin{equation*}
\left(\tfrac{3}{2}d-\sum n_i\right)
\left(1 + \frac{D_1-D_3}{-p^2}\right)
\end{equation*}
inserted into the integrand.
On the other hand, explicit differentiation in $p$ gives
\begin{align*}
&d + \frac{n_1}{D_1} 2(k_1+p)^2 + \frac{n_2}{D_2} 2(k_2+p)\cdot(k_1+p)\,,\\
&2(k_2+p)\cdot(k_1+p) = D_5-D_1-D_2\,.
\end{align*}
Therefore, we obtain the Larin's relation~\cite{Larin}
\begin{equation}
\Bigl[\tfrac{1}{2}d+n_1-n_3-n_4-n_5
+ \left(\tfrac{3}{2}d-\sum n_i\right) (\mathbf{1}^- - \mathbf{3}^-)
+ n_2 \mathbf{2}^+ (\mathbf{1}^- - \mathbf{5}^-)\Bigr] G = 0
\label{p2:Larin}
\end{equation}
(three more relations follow from the symmetries).
This relation can be used instead of~(\ref{p2:tri2})
to reduce all integrals in the positive sector
to the master integrals~(\ref{p2:masters}), see Fig.~\ref{F:IBP}.
It is surely some linear combination of the IBP relations,
but I have no explicit proof of this fact.

\begin{figure}[ht]
\begin{center}
\begin{picture}(42,42)
\put(21,21){\makebox(0,0){\includegraphics{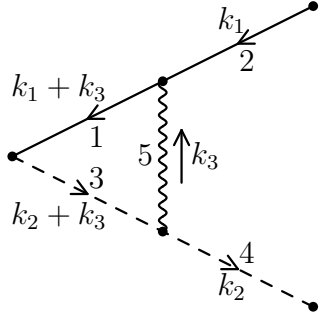}}}
\put(30,39){\makebox(0,0){$k_1$}}
\put(32,33.5){\makebox(0,0){2}}
\put(30,3.5){\makebox(0,0){$k_2$}}
\put(32,8){\makebox(0,0){4}}
\put(7,30){\makebox(0,0){$k_1+k_3$}}
\put(12,23.5){\makebox(0,0){1}}
\put(7,13){\makebox(0,0){$k_2+k_3$}}
\put(12,18){\makebox(0,0){3}}
\put(26.5,21){\makebox(0,0){$k_3$}}
\put(18.5,21){\makebox(0,0){5}}
\end{picture}
\end{center}
\caption{Triangle subdiagram}
\label{F:tri}
\end{figure}

The IBP relations~(\ref{p2:tri1}), (\ref{p2:tri2}) are particular cases
of the triangle relation~\cite{CT:81}.
Suppose a diagram we are considering contains a subdiagram in Fig.~\ref{F:tri}:
\begin{align*}
&D_1 = m_1^2 - (k_1+k_3)^2\,,\qquad
D_2 = m_1^2 - k_1^2\,,\\
&D_3 = m_2^2 - (k_2+k_3)^2\,,\qquad
D_4 = m_2^2 - k_2^2\,,\qquad
D_5 = - k_3^2
\end{align*}
(any number of lines of any kinds may be attached to the left vertex;
however, it is essential that only one line is attached
to each of the two right vertices of the triangle).
The $\partial_3\cdot k_3$ IBP relation is
\begin{equation}
\left[ d - n_1 - n_3 - 2 n_5
+ n_1 \mathbf{1}^+ (\mathbf{2}^- - \mathbf{5}^-)
+ n_3 \mathbf{3}^+ (\mathbf{4}^- - \mathbf{5}^-)
\right] I = 0
\label{p2:tri}
\end{equation}
(e.\,g., (\ref{p2:tri2})).
A single application of the triangle relation~(\ref{p2:tri})
reduces $n_2+n_4+n_5$ by 1, and this always allows one
to kill one of the lines 2, 4, 5.
If the line 2 is external instead of internal,
the lowering operator $\mathbf{2}^-$ becomes just the factor $(m_1^2-k_1^2)$
(where $k_1$ is now an external momentum),
and there is no subset of $n$'s whose sum is reduced
(e.\,g., (\ref{p2:tri1})).

There are 3 generic topologies of 3-loop massless propagator diagrams:
\begin{equation}
\raisebox{-11.05mm}{\includegraphics{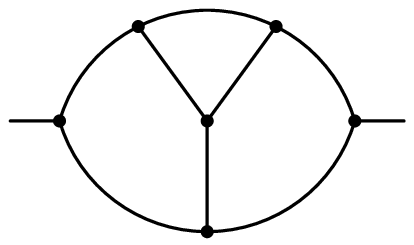}}\,,\qquad
\raisebox{-11.05mm}{\includegraphics{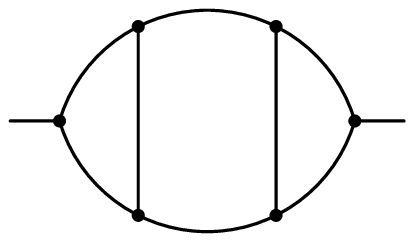}}\,,\qquad
\raisebox{-11.05mm}{\includegraphics{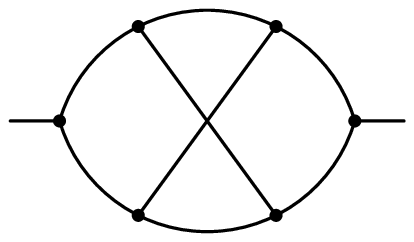}}\,.
\label{p2:Q3t}
\end{equation}
Each has 8 denominators.
There are 9 scalar products of 3 loop momenta $k_i$ and the external momentum $p$.
Therefore, for each topology, all scalar products in the numerator
can be expressed via the denominators and one selected scalar product.
IBP recurrence relations for these diagrams have been investigated in~\cite{CT:81}.
They can be used to reduce all integrals~(\ref{p2:Q3t}),
with arbitrary integer powers of denominators
and arbitrary (non-negative) powers of the selected scalar product in the numerators,
to linear combinations of 6 master integrals
\begin{equation}
\begin{split}
&\raisebox{-4.0mm}{\includegraphics{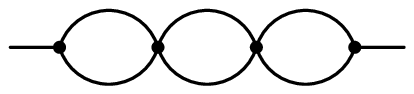}} = G_1^3\,,\qquad
\raisebox{-7.4mm}{\includegraphics{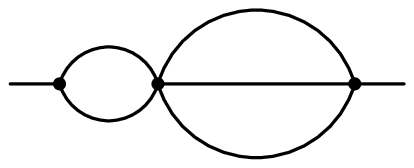}} = G_1 G_2\,,\\
&\raisebox{-11.4mm}{\includegraphics{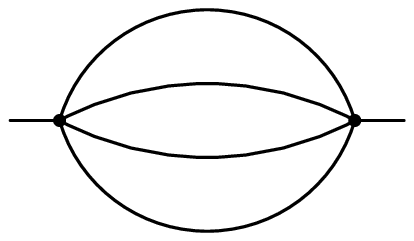}} = G_3\,,\qquad
\raisebox{-11.4mm}{\includegraphics{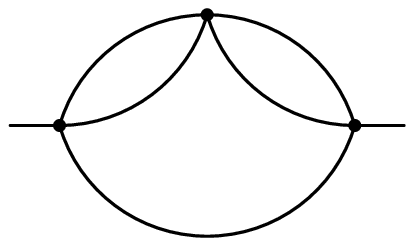}} \sim \frac{G_1^2}{G_2} G_3\,,\\
&\raisebox{-11.4mm}{\includegraphics{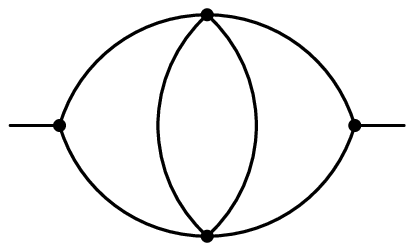}}
= G_1 G(1,1,1,1,2-\tfrac{d}{2})\,,\qquad
\raisebox{-11.4mm}{\includegraphics{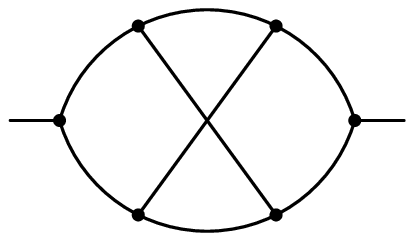}}\,.
\end{split}
\label{p2:Q3b}
\end{equation}
This algorithm has been implemented in the \texttt{SCHOONSCHIP}~\cite{SCHOONSCHIP}
package \texttt{Mincer}~\cite{Mincer:1} and later re-implemented~\cite{Mincer:2}
in \texttt{FORM}~\cite{FORM}%
\footnote{Unfortunately, \texttt{Mincer} does not produce linear combinations
of 6 master integrals~(\ref{p2:Q3b});
recursively 1-loop integrals are expressed via $\Gamma$ functions
and expanded in $\varepsilon=2-d/2$,
so that the contributions from the first 4 master integrals cannot be separated.}.
It has also been implemented in the \texttt{REDUCE}~\cite{REDUCE,G:97}
package \texttt{Slicer}~\cite{Slicer}.
Only the last, non-planar, topology in~(\ref{p2:Q3t})
involves the last, non-planar, master integral in~(\ref{p2:Q3b}).

The first 4 master integrals are trivial: they are expressed
via $G_n$~(\ref{p2:Gn}), and hence via $\Gamma$-functions.
The 4-th one differs from the 3-rd one ($G_3$)
by replacing the two-loop subdiagram:
the second one in~(\ref{p2:masters}) ($G_2/(-k^2)^{3-d}$)
by the first one ($G_1^2/(-k^2)^{4-d}$).
Therefore, it can be obtained from $G_3$ by multiplying by
\begin{equation*}
\frac{G_1^2 G(1,4-d)}{G_2 G(1,3-d)} = \frac{2d-5}{d-3} \frac{G_1^2}{G_2}\,.
\end{equation*}
The 5-th master integral is proportional to the two-loop diagram
$G(1,1,1,1,n)$ with a non-integer index of the middle line $n=2-d/2$.
The 6-th one, non-planar, is truly three-loop and most difficult.

\section{HQET self-energy diagrams}
\label{S:HQET}

There are 2 generic topologies of 2-loop HQET self-energy diagrams:
\begin{equation}
\raisebox{0.2mm}{\includegraphics{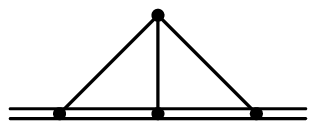}}\,,\qquad
\raisebox{-7.3mm}{\includegraphics{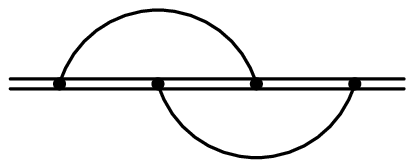}}\,.
\label{p2:H2t}
\end{equation}
The denominators of the second one are linearly dependent,
and there is one irreducible scalar product in the numerator.
All these integrals, with any powers of denominators
(and with any power of the numerator of the second diagram),
can be reduced~\cite{BG:91} (see also~\cite{G:00}) to 2 master integrals
\begin{equation}
\raisebox{-3.825mm}{\includegraphics{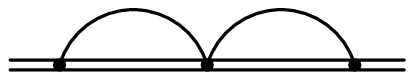}} = I_1^2\,,\quad
\raisebox{-4.2mm}{\includegraphics{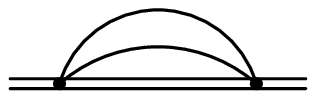}} = I_2\,,
\label{p2:Hb2}
\end{equation}
where
\begin{equation}
I_n =
\raisebox{-8mm}{\begin{picture}(42,17)
\put(21,8.5){\makebox(0,0){\includegraphics{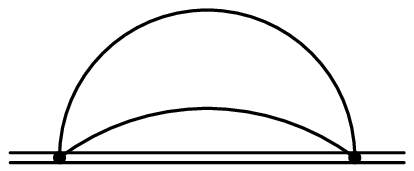}}}
\put(21,11){\makebox(0,0){{$\cdots$}}}
\end{picture}} =
\Gamma(2n+1-nd) \Gamma^n\left({\textstyle\frac{d}{2}}-1\right)
\label{p2:In}
\end{equation}
is the $n$-loop HQET sunset integral.

There are 10 generic topologies of 3-loop HQET self-energy diagrams:
\begin{equation}
\begin{split}
&\raisebox{0.2mm}{\includegraphics{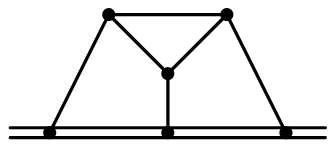}}\,,\qquad
\raisebox{0.2mm}{\includegraphics{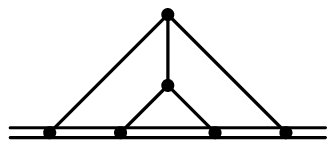}}\,,\\
&\raisebox{0.2mm}{\includegraphics{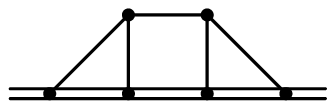}}\,,\qquad
\raisebox{0.2mm}{\includegraphics{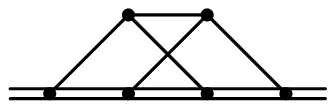}}\,,\\
&\raisebox{-3.4mm}{\includegraphics{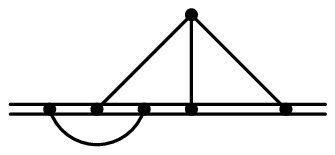}}\,,\qquad
\raisebox{-3.4mm}{\includegraphics{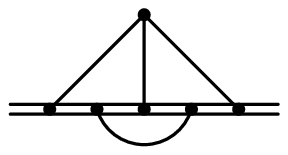}}\,,\qquad
\raisebox{-7mm}{\includegraphics{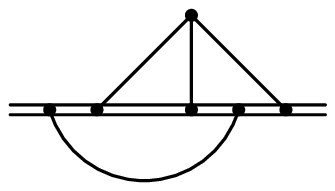}}\,,\\
&\raisebox{-3.4mm}{\includegraphics{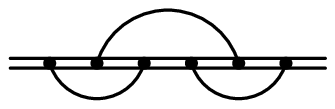}}\,,\qquad
\raisebox{-5.2mm}{\includegraphics{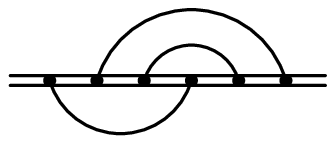}}\,,\qquad
\raisebox{0.2mm}{\includegraphics{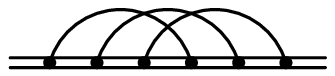}}\,.
\end{split}
\label{p2:H3t}
\end{equation}
Diagrams in the first two rows have one scalar product
which cannot be expressed via denominators;
those in the third row have one linear relation among heavy denominators,
and hence two independent scalar products in the numerator;
those in the last row have two relations among heavy denominators,
and hence three independent scalar products in the numerator.
All these integrals, with any powers of denominators
and irreducible numerators, can be reduced~\cite{G:00}
to 8 master integrals
\begin{equation}
\begin{split}
&\raisebox{-0.25mm}{\includegraphics{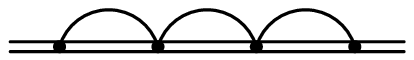}} = I_1^3\,,\qquad
\raisebox{-0.25mm}{\includegraphics{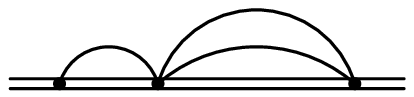}} = I_1 I_2\,,\\
&\raisebox{-0.25mm}{\includegraphics{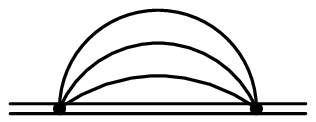}} = I_3\,,\qquad
\raisebox{-0.25mm}{\includegraphics{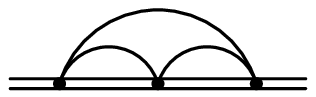}} \sim \frac{I_1^2}{I_2} I_3\,,\\
&\raisebox{-0.25mm}{\includegraphics{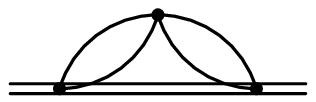}} \sim \frac{G_1^2}{G_2} I_3\,,\qquad
\raisebox{-0.25mm}{\includegraphics{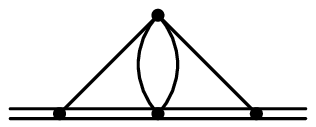}} = G_1 I(1,1,1,1,2-\tfrac{d}{2})\,,\\
&\raisebox{-8.25mm}{\includegraphics{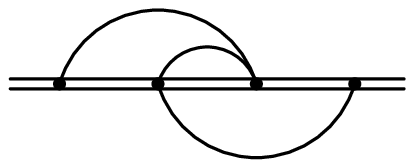}} = I_1 J(1,1,3-d,1,1)\,,\qquad
\raisebox{-0.25mm}{\includegraphics{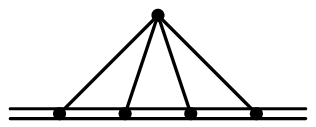}}\,.
\end{split}
\label{p2:H3b}
\end{equation}
using integration by parts.
This reduction algorithm has been implemented
as a \texttt{REDUCE} package \texttt{Grinder}~\cite{G:00}.
The first 5 master integrals can be easily expressed
via $\Gamma$ functions, exactly in $d$ dimensions.
The next 2 ones reduce to two-loop ones with a single
$d$-dependent index; the last one is truly three-loop.

\section{Equivalence of IBP relations for integrals with the\\
same total number of loop and external momenta}
\label{S:Equiv}

IBP relations for integrals with the same $M=L+E$ but different $E$
can be made equivalent~\cite{BS:00}
(then they differ only by boundary conditions --- sets of trivial sectors).
In other words, the IBP relations for an $L$-loop Feynman integral~(\ref{Intro:I})
with $E$ external momenta can be reduced (after some re-definitions)
to exactly the same form~(\ref{IBP:k}) as for the $M$-loop vacuum integral
(in which we also integrate in $d^d p_1\cdots d^d p_E$).

Consider some $L$-loop integral $I(n_1,\ldots,n_N)$
with $E$ external momenta
at some kinematic point $s_{ij}=s^0_{ij}$ ($i,j>L$, $j\ge i$).
We number all such $(i,j)$ pairs by integers $a\in[N+1,K]$
in some way, where
\begin{equation}
K = N + N_E = \frac{M(M+1)}{2}
\label{Equiv:NE}
\end{equation}
is the number of all scalar products $s_{ij}$
of $M$ vectors $q_i$ ($j\ge i$),
and introduce the quantities $D_a=-s_{ij}+s^0_{ij}=-s_{ij}+m_a^2$
($a\in[N+1,K]$) for these $(i,j)$ pairs.
Now all the quantities $D_a$ can be written in a uniform way~(\ref{Intro:Ds})
(with both sums up to $M$).
All the scalar products $s_{ij}$ can be expressed via $D_a$
in a uniform way~(\ref{Intro:sD}) with the sum up to $K$.

Let's expand this integral $I(n_1,\ldots,n_N)$~(\ref{Intro:I})
in a formal series in $s_{ij}-s^0_{ij}$ ($L<i\le j\le M$):
\begin{equation}
I(n_1,\ldots,n_N) = \sum_{n_{N+1}=1}^\infty \cdots \sum_{n_K=1}^\infty
I(n_1,\ldots,n_N,n_{N+1},\ldots,n_K)
D_{N+1}^{n_{N+1}-1} \cdots D_K^{n_K-1}\,.
\label{Equiv:exp}
\end{equation}
In fact, we are mainly interested in the value of our integral at this kinematic point
($n_{N+1}=\cdots=n_K=1$),
the derivatives (given by higher values of these indices)
are not our primary goal.
We assume $I=0$ if any index $n_{N+1}$, \dots, $n_K$ is $\le0$,
so that there is only one sector with respect to each of these indices.

Applying $N$ operators $O_{ij}$ with $i\in[1,L]$, $j\ge i$ to the integrand $f$, we get
the usual IBP relations~(\ref{IBP:k}).
In other words, the ``evolution'' of $I(n_1,\ldots,n_N,n_{N+1},\ldots,n_K)$
with respect to the ``internal'' indices $n_1$, \dots, $n_N$
is governed by $N$ ``vacuum'' IBP relations~(\ref{IBP:k}).

We need $N_E$ additional relations which govern the ``evolution''
with respect to the ``external'' indices $n_{N+1}$, \dots, $n_K$.
To derive them, we apply $N_E$ operators $O_{ij}=q_j\cdot \partial_i$
($L<i\le j\le M$) to the definition~(\ref{Equiv:exp}).
In the left-hand side, we get
\begin{equation*}
- \sum_{a=1}^N \sum_{b=1}^N \sum_{m=1}^L A_a^{mi} A^b_{mj}
\hat{\mathbf{a}}^+ \left(\mathbf{b}^- - m_b^2\right)
I(n_1,\ldots,n_N)\,;
\end{equation*}
here each integral can be expanded like~(\ref{Equiv:exp}).
In the right-hand side, $O_{ij}$ acts on the product of $D_a$ with $a\in[N+1,K]$.
To get the coefficient of $D_{N+1}^{n_{N+1}-1} \cdots D_K^{n_K-1}$,
we shift the index $n_a\to n_a+1$ in terms with $1/D_a$,
and $n_b\to n_b-1$ in terms with $D_b$, and obtain
\begin{equation*}
\sum_{a=N+1}^K \sum_{b=N+1}^K \sum_{m=L+1}^M A_a^{mi} A^b_{mj} (1+\delta_{mi})
\left(\mathbf{b}^- - m_b^2\right) \hat{\mathbf{a}}^+
I(n_1,\ldots,n_N,n_{N+1},\ldots,n_K)
\end{equation*}
(note the opposite order of the operators).
In order to combine this with the left-hand side,
we commute the operators~(\ref{IBP:comm3}) and use
\begin{equation*}
\sum_a A_a^{ij} A^a_{i'j'} = \delta^i_{i'} \delta^j_{j'}\qquad
(j\ge i,j'\ge i')\,.
\end{equation*}
Finally, the relations for the ``external'' indices are
\begin{equation}
\begin{split}
&\left[ (E+1) \delta_{ij}
- \sum_{a=1}^K \sum_{b=1}^K \sum_{m=1}^M A_a^{mi} A^b_{mj} (1+\delta_{mi})
\hat{\mathbf{a}}^+ \left(\mathbf{b}^- - m_b^2\right) \right]\\
&\qquad{}\times I(n_1,\ldots,n_N,n_{N+1},\ldots,n_K) = 0\,.
\end{split}
\label{Equiv:ext}
\end{equation}
They are similar to the ``internal'' ones~(\ref{IBP:k}),
but contain $E+1$ instead of $d$.

\begin{sloppypar}
We can make these relation exactly ``vacuum'' ones using the substitution
\begin{equation}
\begin{split}
&I(n_1,\ldots,n_N) = D^{(E+1-d)/2} \tilde{I}(n_1,\ldots,n_N)\,,\\
&D = \frac{\det s_{ij}}{\det s^0_{ij}}\,,\qquad
i,j\in[L+1,M]\,.
\end{split}
\label{Equiv:D}
\end{equation}
The new functions $\tilde{I}(n_1,\ldots,n_N)$ are expanded in $D_a$ with $a>N$
in the same way as in~(\ref{Equiv:exp}),
with the coefficients $\tilde{I}(n_1,\ldots,n_N,n_{N+1},\ldots,n_K)$.
At $s_{ij}=s^0_{ij}$ $\tilde{I}$ coincides with $I$:
$\tilde{I}(n_1,\ldots,n_N,1,\ldots,1)=I(n_1,\ldots,n_N,1,\ldots,1)$;
higher $\tilde{I}(n_1,\ldots,n_N,n_{N+1},\ldots,n_K)$ are linear combinations
of $I(n_1,\ldots,n_N,n_{N+1},\ldots,n_K)$.
The operator $O_{ij}$ can be written as
\begin{equation*}
O_{ij} = \sum_{m=L+1}^M (1+\delta_{mi}) s_{mj} \partial_{mi}\,,\qquad
\partial_{mi} = \frac{\partial}{\partial s_{mi}}\,.
\end{equation*}
The derivatives of $D$ are
\begin{equation*}
\partial_{ij} D = (2-\delta_{ij}) D \cdot (s^{-1})_{ji}\qquad
(i,j\in[L+1,M])
\end{equation*}
(the extra factor 2 at $i\ne j$ comes from the fact that $s_{ij}$ appears
in the determinant $D$ twice, at the positions $ij$ and $ji$).
Therefore,
\begin{equation*}
O_{ij} D^{(E+1-d)/2} = (E+1-d) D^{(E+1-d)/2} \delta_{ij}\,,
\end{equation*}
and an additional term $(d-E-1) \delta_{ij}$ appears in~(\ref{Equiv:ext}).
All relations for $\tilde{I}(n_1,\ldots,n_N,n_{N+1},\ldots,n_K)$
have the same form as for the $M=L+E$ loop vacuum integral $I_0(n_1,\ldots,n_K)$.
\end{sloppypar}

The boundary conditions may differ.
In the case of $\tilde{I}(n_1,\ldots,n_N,n_{N+1},\ldots,n_K)$,
all sectors with $n_a\le0$ ($a\in[N+1,K]$) are trivial;
this does not have to be so for the corresponding vacuum diagram.

Now we shall consider a few examples.

\begin{figure}[ht]
\begin{center}
\begin{picture}(74,36)
\put(19,21.25){\makebox(0,0){\includegraphics{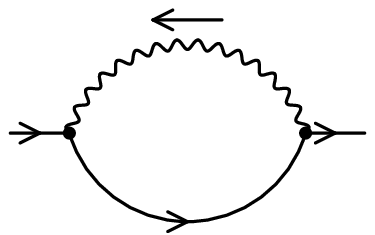}}}
\put(19,10){\makebox(0,0)[t]{$k+p$}}
\put(19,32.5){\makebox(0,0)[b]{$k$}}
\put(19,12){\makebox(0,0)[b]{1}}
\put(19,28){\makebox(0,0)[t]{2}}
\put(4,19){\makebox(0,0)[t]{$p$}}
\put(34,19){\makebox(0,0)[t]{$p$}}
\put(19,0){\makebox(0,0)[b]{a}}
\put(61,20){\makebox(0,0){\includegraphics{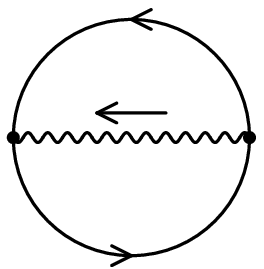}}}
\put(61,7){\makebox(0,0)[t]{$k_1+k_2$}}
\put(61,32){\makebox(0,0)[b]{$k_2$}}
\put(61,23.5){\makebox(0,0)[b]{$k_1$}}
\put(61,9){\makebox(0,0)[b]{1}}
\put(61,31){\makebox(0,0)[t]{3}}
\put(61,19){\makebox(0,0)[t]{2}}
\put(61,0){\makebox(0,0)[b]{b}}
\end{picture}
\end{center}
\caption{Massive 1-loop self-energy and 2-loop vacuum diagram}
\label{F:m1}
\end{figure}

Let's consider the 1-loop self-energy diagram
(Fig.~\ref{F:m1}a; we set $m=1$):
\begin{equation}
M(n_1,n_2) = \frac{1}{i\pi^{d/2}}
\int \frac{d^d k}{D_1^{n_1} D_2^{n_2}}\,,\qquad
D_1 = 1 - (k+p)^2\,,\qquad
D_2 = -k^2\,.
\label{Equiv:m1}
\end{equation}
This integral vanishes at $n_1\le0$.
Suppose we want to calculate it on the mass shell $p^2=1$.
We introduce $D_3=1-p^2$,
and re-express the integral as
\begin{equation}
M(n_1,n_2) = (p^2)^{(2-d)/2} \tilde{M}(n_1,n_2)\,,
\label{Equiv:m1D}
\end{equation}
according to~(\ref{Equiv:D}).
Then we expand it in $D_3$:
\begin{equation}
\tilde{M}(n_1,n_2) = \sum_{n_3=1}^\infty \tilde{M}(n_1,n_2,n_3) D_3^{n_3-1}
\label{Equiv:m1exp}
\end{equation}
($\tilde{M}(n_1,n_2,n_3)$ vanishes at $n_3\le0$).
There is 1 master integral:
\begin{equation}
\tilde{M}(n_1,n_2,n_3) = c(n_1,n_2,n_3) \tilde{M}(1,0,1)\,.
\label{Equiv:Mmaster}
\end{equation}

The $\partial\cdot k$ and $\partial\cdot(k+p)$ IBP relations for $M(n_1,n_2)$ are
\begin{equation*}
\begin{split}
&\left[ d - \mathbf{n}_1 - 2 \mathbf{n}_2
+ \mathbf{n}_1 \mathbf{1}^+ \left( D_3 - \mathbf{2}^- \right) \right]
M(n_1,n_2) = 0\,,\\
&\left[ d - 2 \mathbf{n}_1 - \mathbf{n}_2
+ 2 \mathbf{n}_1 \mathbf{1}^+
+ \mathbf{n}_2 \mathbf{2}^+ \left( D_3 - \mathbf{1}^- \right) \right]
M(n_1,n_2) = 0
\end{split}
\end{equation*}
(here $D_3$ is an external kinematic parameter).
For the expansion coefficients this means
\begin{equation}
\begin{split}
&\left[ d - \mathbf{n}_1 - 2 \mathbf{n}_2
+ \mathbf{n}_1 \mathbf{1}^+ \left( \mathbf{3}^- - \mathbf{2}^- \right) \right]
\tilde{M}(n_1,n_2,n_3) = 0\,,\\
&\left[ d - 2 \mathbf{n}_1 - \mathbf{n}_2
+ 2 \mathbf{n}_1 \mathbf{1}^+
+ \mathbf{n}_2 \mathbf{2}^+ \left( \mathbf{3}^- - \mathbf{1}^- \right) \right]
\tilde{M}(n_1,n_2,n_3) = 0
\end{split}
\label{Equiv:IBPMk}
\end{equation}
(we shift the summation index $n_3\to n_3-1$ in the terms with $D_3$).
In order to find the ``evolution'' in $n_3$,
we apply $p\cdot\partial/\partial p$ to~(\ref{Equiv:m1exp}):
\begin{align*}
&p \cdot \frac{\partial}{\partial p} M(n_1,n_2) =
\left[ - \mathbf{n}_1 + \mathbf{n}_1 \mathbf{1}^+
\left( 2 + \mathbf{2}^- - D_3 \right) \right]
M(n_1,n_2)\,,\\
&p \cdot \frac{\partial}{\partial p} (p^2)^{(d-2)/2} =
(d-2) (p^2)^{(d-2)/2}\,,\\
&\sum_{n_3=1}^\infty \tilde{M}(n_1,n_2,n_3)
p \cdot \frac{\partial}{\partial p} D_3^{n_3-1} =
2 \sum_{n_3=1}^\infty \tilde{M}(n_1,n_2,n_3)
(n_3-1) \left( D_3^{n_3-1} - D_3^{n_3-2} \right)\\
&\qquad{} = 2 \sum_{n_3=1}^\infty
\left[ \left( \mathbf{n}_3 - 1 - \mathbf{n}_3 \mathbf{3}^+ \right)
\tilde{M}(n_1,n_2,n_3) \right] D_3^{n_3-1}\,.
\end{align*}
Collecting these pieces together, we get
\begin{equation}
\left[ d - \mathbf{n}_1 - 2 \mathbf{n}_3
+ \mathbf{n}_1 \mathbf{1}^+ \left( 2 + \mathbf{2}^- - \mathbf{3}^- \right)
+ 2 \mathbf{n}_3 \mathbf{3}^+ \right]
\tilde{M}(n_1,n_2,n_3) = 0\,.
\label{Equiv:IBPMp}
\end{equation}

Now let's consider the 2-loop vacuum diagram (Fig.~\ref{F:m1}b, $m=1$):
\begin{equation}
\begin{split}
&V(n_1,n_2,n_3) = \frac{1}{(i\pi^{d/2})^2}
\int \frac{d^d k_1 d^d k_2}{D_1^{n_1} D_2^{n_2} D_3^{n_3}}\,,\\
&D_1 = 1 - (k_1+k_2)^2\,,\qquad
D_2 = - k_1^2\,,\qquad
D_3 = 1 - k_2^2\,.
\end{split}
\label{Equiv:v2}
\end{equation}
This integral vanishes at $n_1\le0$ or $n_3\le0$.
The $\partial_1\cdot k_1$, $\partial_1\cdot(k_1-k_2)$,
and $\partial_2\cdot k_2$ IBP relations for $V(n_1,n_2,n_3)$
have exactly the same form as~(\ref{Equiv:IBPMk}), (\ref{Equiv:IBPMp}),
as expected.
There is 1 master integral:
\begin{equation}
V(n_1,n_2,n_3) = c_0(n_1,n_2,n_3) V(1,0,1)\,.
\label{Equiv:Vmaster}
\end{equation}

In this problem, the boundary conditions for $\tilde{M}(n_1,n_2,n_3)$
and $V(n_1,n_2,n_3)$ coincide, and hence $c(n_1,n_2,n_3)=c_0(n_1,n_2,n_3)$.
Therefore, the 1-loop on-shell self-energy diagram $M(n_1,n_2)=\tilde{M}(n_1,n_2,1)$
is related to the 2-loop vacuum diagram $V(n_1,n_2,1)$
(in which the index of the ``former external'' line is $n_3=1$):
\begin{equation}
\frac{M(n_1,n_2)}{M(1,0)} = \frac{V(n_1,n_2,1)}{V(1,0,1)}\,.
\label{Equiv:MV}
\end{equation}
Explicit expressions for $M(n_1,n_2)$ and $V(n_1,n_2,n_3)$
can be found, e.\,g., in the textbook~\cite{G:07}.
It is easy to check the relation~(\ref{Equiv:MV})
using
\begin{equation*}
\Gamma(z) \Gamma(1-z) = \frac{\pi}{\sin \pi z}\,.
\end{equation*}

Reduction of 2-loop on-shell self-energy diagrams to master integrals
has been considered in~\cite{B:92}:
\begin{align}
\raisebox{-6.2mm}{\includegraphics{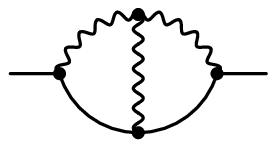}}
={}& c_1(n_1,\ldots,n_5) \raisebox{-9.2mm}{\includegraphics{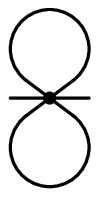}}
+ c_2(n_1,\ldots,n_5) \raisebox{-6.2mm}{\includegraphics{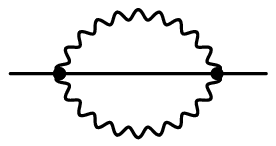}}\,,
\label{Equiv:M2}\\
\raisebox{-6.2mm}{\includegraphics{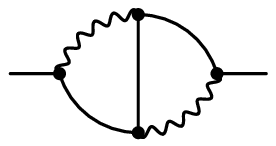}}
={}& c_3(n_1,\ldots,n_5) \raisebox{-9.2mm}{\includegraphics{m2b1.eps}}
+ c_4(n_1,\ldots,n_5) \raisebox{-6.2mm}{\includegraphics{m2b2.eps}}
\nonumber\\
&{} + c_5(n_1,\ldots,n_5) \raisebox{-6.2mm}{\includegraphics{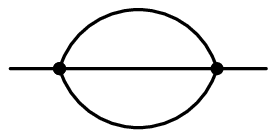}}\,.
\label{Equiv:N2}
\end{align}
These IBP reduction relations are equivalent to those for 3-loop vacuum diagrams
(also considered in~\cite{B:92}):
\begin{align}
\raisebox{-7.2mm}{\includegraphics{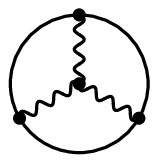}}
={}& c'_1(n_1,\ldots,n_6) \raisebox{-9.7mm}{\includegraphics{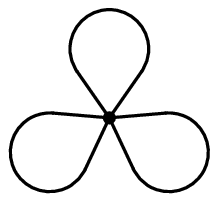}}
+ c'_2(n_1,\ldots,n_6) \raisebox{-7.2mm}{\includegraphics{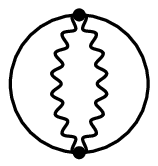}}\,,
\label{Equiv:VM}\\
\raisebox{-7.2mm}{\includegraphics{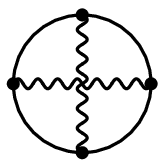}}
={}& c'_3(n_1,\ldots,n_6) \raisebox{-9.7mm}{\includegraphics{v3b1.eps}}
+ c'_4(n_1,\ldots,n_6) \raisebox{-7.2mm}{\includegraphics{v3b2.eps}}
\nonumber\\
&{} +  c'_5(n_1,\ldots,n_6) \raisebox{-7.2mm}{\includegraphics{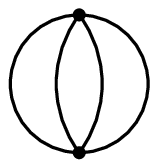}}\,.
\label{Equiv:VN}
\end{align}
However, boundary conditions differ:
on-shell integrals vanish at $n_6\le0$,
but vacuum integrals don't.
Therefore, it is not possible to obtain $c_k(n_1,\ldots,n_5)$
from $c'_k(n_1,\ldots,n_5,1)$.
Reduction for both classes of diagrams is implemented
in a \texttt{REDUCE} package \texttt{Recursor}~\cite{B:92}.
For 2-loop on-shell diagrams, $n_6=1$ is specified,
and a special control flag is set
which is responsible for nullifying all integrals with $n_6\le0$
during reduction%
\footnote{The redefinition~(\ref{Equiv:D}) is not used in this package;
therefore, some terms in the vacuum IBP relations are absent in the on-shell ones.
Their inclusion is controlled by the same flag.}.

\begin{figure}[ht]
\begin{center}
\begin{picture}(42.8,48.8)
\put(21.4,24.4){\makebox(0,0){\includegraphics{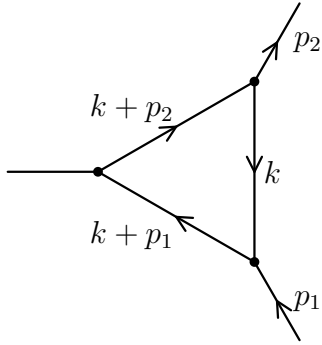}}}
\put(36,24.4){\makebox(0,0)[l]{$k$}}
\put(40,7.2){\makebox(0,0)[l]{$p_1$}}
\put(40,41.6){\makebox(0,0)[l]{$p_2$}}
\put(24,31){\makebox(0,0)[br]{$k+p_2$}}
\put(24,17.8){\makebox(0,0)[tr]{$k+p_1$}}
\end{picture}
\end{center}
\caption{1-loop vertex diagram (all lines are massless)}
\label{F:t1}
\end{figure}

Our next example is the 1-loop massless vertex diagram (Fig.~\ref{F:t1})
with 2 on-shell legs: $p_1^2=p_2^2=0$.
Its IBP relations are equivalent to those for the 2-loop massless
self-energy diagram (Fig.~\ref{F:Q2}),
because both are equivalent to the same 3-loop vacuum diagram
with 1 massive line.
These 2-loop self-energy diagrams are expressed via 2 master integrals
(Sect.~\ref{S:p2}):
\begin{equation}
\begin{split}
\raisebox{-7.6mm}{\begin{picture}(42,17)
\put(21,8.5){\makebox(0,0){\includegraphics{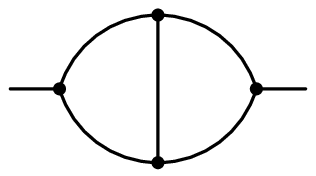}}}
\put(11,14.5){\makebox(0,0){$n_2$}}
\put(11,2.5){\makebox(0,0){$n_1$}}
\put(31,14.5){\makebox(0,0){$n_5$}}
\put(31,2.5){\makebox(0,0){$n_4$}}
\put(24,8.5){\makebox(0,0){$n_3$}}
\end{picture}}
={}& c_1(n_1,\ldots,n_5) \raisebox{-7.6mm}{\includegraphics{q2b1.eps}}\\
&{}+ c_2(n_1,\ldots,n_5) \raisebox{-7.6mm}{\includegraphics{q2b2.eps}}\,.
\end{split}
\label{Equiv:q2}
\end{equation}
The boundary conditions for the 1-loop vertex are different:
integrals with $n_{4,5}\le0$ vanish.
Such integrals produce the second master integral in~(\ref{Equiv:q2}).
This leads to a very simple prescription:
replace the second master integral in~(\ref{Equiv:q2}) by $0$
and the first one by the vertex master integral,
\begin{equation}
\raisebox{-11.3mm}{\begin{picture}(23,24.4)
\put(10.7,12.2){\makebox(0,0){\includegraphics{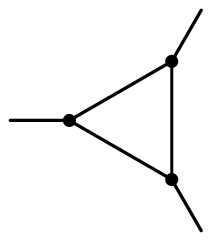}}}
\put(20,12.2){\makebox(0,0){$n_3$}}
\put(11,18.2){\makebox(0,0){$n_2$}}
\put(11,6.2){\makebox(0,0){$n_1$}}
\end{picture}}
= c_1(n_1,n_2,n_3,1,1) \raisebox{-7.6mm}{\includegraphics{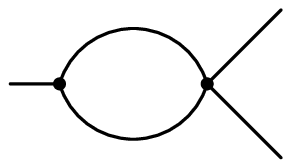}}\,.
\label{Equiv:t1}
\end{equation}

Similarly, reduction for 2-loop massless vertex diagrams with $p_1^2=p_2^2=0$
is equivalent to 3-loop massless self-energy diagrams~\cite{BS:00}.
This method is applicable to diagrams with any number of external legs.

\section{Gr\"obner-bases methods}
\label{S:SBases}

For complicated problems, constructing reduction algorithms by hand
becomes impractical, and it is desirable to automate this process.
Some of such systematic approaches are based on Gr\"obner bases.
A simple introduction to Gr\"obner bases for systems of polynomial equations
with commuting variables is given in Appendix~\ref{S:Groebner}.
In the IBP reduction problem, polynomials of non-commuting operators are used.
An approach using the shift operators~(\ref{IBP:Oper}) was proposed in~\cite{SS:06,S:06}
and implemented in a \texttt{\textit{Mathematica}} program \texttt{FIRE}~\cite{FIRE}.

\begin{figure}[ht]
\begin{center}
\begin{picture}(76,76)
\put(37,37){\makebox(0,0){\includegraphics{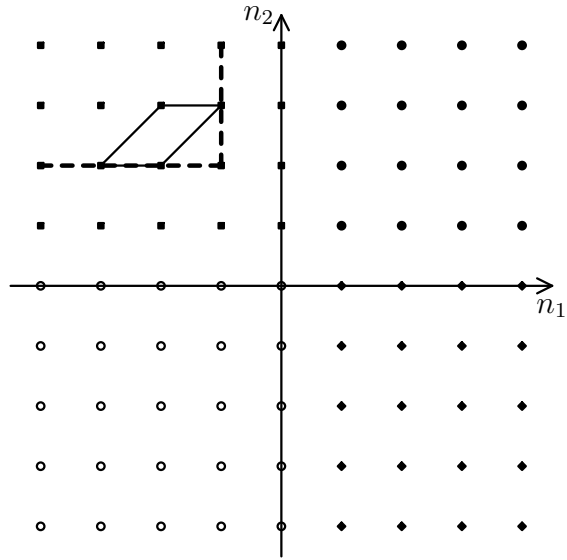}}}
\put(73,34){\makebox(0,0){$n_1$}}
\put(34,73){\makebox(0,0){$n_2$}}
\end{picture}
\end{center}
\caption{Normal form of IBP relations in a sector}
\label{F:Normal}
\end{figure}

Suppose we are considering the sector $n_1\le0$, $n_2>0$ in Fig.~\ref{F:Normal}.
Any integral in the sector can be expressed via the integral at its corner
using the shift operators $\mathbf{1}^-$, $\mathbf{2}^+$:
\begin{equation}
I(n_1,n_2) = \left(\mathbf{1}^-\right)^{-n_1} \left(\mathbf{2}^+\right)^{n_2-1} I(0,1)\,.
\label{SBases:corner}
\end{equation}

All IBP relations in this sector can be written in the normal form
containing only $\mathbf{1}^-$ and $\mathbf{2}^+$:
we act on the original form of these relations by $\mathbf{1}^-$ and $\mathbf{2}^+$
sufficiently many times to get rid of $\mathbf{1}^+$, $\mathbf{2}^-$
(but not too many times, see Fig.~\ref{F:Normal}).
They are
\begin{equation}
\sum_{j_1,j_2\ge0} C_{j_1 j_2}(\mathbf{n}_a)
\left(\mathbf{1}^-\right)^{j_1} \left(\mathbf{2}^+\right)^{j_1}
\sim 0
\label{SBases:Normal}
\end{equation}
(this means that these operator polynomials give 0 when applied to $I$;
note that $\mathbf{n}_a$ don't commute with the shift operators).
We need to fix some total order of the monomials
constructed from $\mathbf{1}^-$ and $\mathbf{2}^+$;
this is equivalent to fixing a shift-invariant total order
for the integrals in our sector (Sect.~\ref{S:IBP}).
Now we can construct the Gr\"obner basis for the system~(\ref{SBases:Normal})
(more exactly $S$-basis~\cite{SS:06,S:06}),
reduce the monomial
$\left(\mathbf{1}^-\right)^{-n_1} \left(\mathbf{2}^+\right)^{n_2-1}$~(\ref{SBases:corner})
with respect to it, and apply it to $I(0,1)$.
Integrals from lower sectors are considered trivial and already known.
Irreducible monomials give master integrals in this sector.

The idea to use Gr\"obner bases for IBP reduction was originally proposed
in~\cite{T:98} in a somewhat different setting.
Let's assume that each line has a separate mass $m_a$,
and consider integrals without numerators
(numerators can be eliminated by shifting $d$~\cite{T:96}).
Differentiation in $m_a^2$ is equivalent to the raising operator~(\ref{IBP:hat}):
\begin{equation}
\frac{\partial}{\partial m_a^2} \Rightarrow - \hat{\mathbf{a}}^+\,.
\label{SBases:dm}
\end{equation}
Any integral in the positive sector can be expressed
via the integral at its corner:
\begin{equation}
I(n_1,\ldots,n_1) =
\left(-\frac{\partial}{\partial m_1^2}\right)^{n_1-1} \cdots
\left(-\frac{\partial}{\partial m_N^2}\right)^{n_N-1} I(1,\ldots,1)\,.
\label{SBases:cornerd}
\end{equation}
The IBP relations can be written as
\begin{equation}
\sum C_{j_1\ldots j_N}(m_1^2,\ldots,m_N^2)
\left(\frac{\partial}{\partial m_1^2}\right)^{j_1} \cdots
\left(\frac{\partial}{\partial m_N^2}\right)^{j_N}
\sim 0\,.
\label{SBases:pde}
\end{equation}
Defining a total order for monomials constructed from $\partial/\partial m_a^2$,
we can find the Gr\"obner basis of the system~(\ref{SBases:pde}).
Then we can reduce the monomial in~(\ref{SBases:cornerd}) with respect to this basis.
Irreducible monomials give the master integrals.

This method has been implemented in \texttt{Maple}
and successfully applied~\cite{T:04} to the problem of reduction
of 2-loop self-energy diagrams with all 5 masses different
and arbitrary $p^2$~\cite{T:97}.
If there are many zero (or equal) masses in the problem,
this approach requires to solve a more difficult problem
with all masses being different first,
and this may lead to very lengthy intermediate expressions.

\section{Baikov's method}
\label{S:Baikov}

The Feynman integral~(\ref{Intro:I}) can be written as an integral
in scalar products $s_{ij}$%
\footnote{Here we consider the Euclidean case, in order not to have complications
with signs and $\pm i0$.}.
The integration measure is
\begin{equation}
d^d k_1 d^d k_2 \cdots d^d k_L =
d^{M-1} k_{1||} d^{d-M+1} k_{1\bot}
d^{M-2} k_{2||} d^{d-M+2} k_{2\bot} \cdots
d^{M-L} k_{L||} d^{d-M+L} k_{L\bot}\,,
\label{Baikov:dk}
\end{equation}
where $k_{1||}$ lies in the subspace spanned by $k_2$, \dots, $k_L$, $p_1$, \dots, $p_E$;
$k_{2||}$ lies in the subspace spanned by $k_3$, \dots, $k_L$, $p_1$, \dots, $p_E$;
and so on.

The volume elements $d^{M-i} k_{i||}$ are
\begin{equation}
\begin{split}
&d^{M-1} k_{1||} =
\frac{d s_{12} d s_{13} \cdots d s_{1M}}{G^{1/2}(k_2,\ldots,k_L,p_1,\ldots,p_E)}\,,\\
&d^{M-2} k_{2||} =
\frac{d s_{23} d s_{24} \cdots d s_{2M}}{G^{1/2}(k_3,\ldots,k_L,p_1,\ldots,p_E)}\,,\\
&\cdots\\
&d^{M-L} k_{L||} =
\frac{d s_{L,L+1} d s_{L,L+2} \cdots d s_{LM}}{G^{1/2}(p_1,\ldots,p_E)}\,,
\end{split}
\label{Baikov:par}
\end{equation}
where
\begin{equation}
G(q_1,\ldots,q_n) = \det q_i\cdot q_j
\label{Baikov:Gram}
\end{equation}
is the Gram determinant
($G^{1/2}(q_1,\ldots,q_n)$ is the volume of the parallelogram
formed by $q_1$, \dots, $q_n$).

We can integrate over the angles in $d^n k_{i\bot}$:
\begin{equation*}
d^n k_{i\bot} = \frac{1}{2} \Omega_n k_{i\bot}^{n-2} d k^2_{i\bot}\,,
\end{equation*}
where
\begin{equation}
\Omega_n = \frac{2 \pi^{n/2}}{\Gamma(n/2)}
\label{Baikov:Omega}
\end{equation}
is the $n$-dimensional full solid angle.
Then we replace $d k^2_{i\bot} = d s_{ii}$;
$k_{i\bot}$ is the height of the parallelogram with the base
formed by $k_{i+1}$, \dots, $k_L$, $p_1$, \dots, $p_E$
and the extra vector $k_i$,
and this is the volume of the whole parallelogram
divided by the area of its base.
Therefore,
\begin{equation}
\begin{split}
&d^{d-M+1} k_{1\bot} = \frac{1}{2} \Omega_{d-M+1}
\left(\frac{G(k_1,\ldots,k_L,p_1,\ldots,p_E)}{G(k_2,\ldots,k_L,p_1,\ldots,p_E)}\right)^{(d-M-1)/2}
d s_{11}\,,\\
&d^{d-M+2} k_{2\bot} = \frac{1}{2} \Omega_{d-M+2}
\left(\frac{G(k_2,\ldots,k_L,p_1,\ldots,p_E)}{G(k_3,\ldots,k_L,p_1,\ldots,p_E)}\right)^{(d-M)/2}
d s_{22}\,,\\
&\cdots\\
&d^{d-M+L} k_{L\bot} = \frac{1}{2} \Omega_{d-M+L}
\left(\frac{G(k_L,p_1,\ldots,p_E)}{G(p_1,\ldots,p_E)}\right)^{(d-M+L-2)/2}
d s_{LL}\,.
\end{split}
\label{Baikov:perp}
\end{equation}

All the Gram determinants except the first and the last ones cancel
in the measure~(\ref{Baikov:dk}), and~\cite{L:10}
\begin{equation}
\begin{split}
&I(n_1,\ldots,n_N) = \frac{1}{\pi^{Ld/2}} \int d^d k_1 \cdots d^d k_L f ={}\\
&\frac{\pi^{-L(L-1)/4-LE/2}}{\prod_{i=1}^L \Gamma\left(\frac{d-M+i}{2}\right)}
G(p_1,\ldots,p_E)^{(-d+E+1)/2}\\
&{}\times \int \prod_{i=1}^L \prod_{j=i}^M d s_{ij}
G(k_1,\ldots,k_L,p_1,\ldots,p_E)^{(d-M-1)/2} f\,.
\end{split}
\label{Baikov:Lee}
\end{equation}
The integration region has a complicated shape;
the Gram determinant vanishes on its boundaries.
This formula can be used as a definition of the $d$-dimensional integral.
Usually, another definition based on the $\alpha$ (or Feynman) parametrization is used.
But then many simple properties, like the possibility to cancel identical brackets
in the numerator and the denominator of the integrand $f$,
become theorems needing non-trivial proofs.
Here the possibility to cancel brackets (depending on $s_{ij}$) is obvious.

The integration variables $x_a=D_a$ ($a\in[1,N]$)~\cite{B:97}
can be used instead of $s_{ij}$:
\begin{equation}
I(n_1,\ldots,n_N) = C \int
\frac{d x_1 \cdots d x_N}{x_1^{n_1} \cdots x_N^{n_N}}
P(x_1-m_1^2,\ldots,x_N-m_N^2)^{(d-M-1)/2}\,,
\label{Baikov:Baikov}
\end{equation}
where the Baikov polynomial is
\begin{equation}
P(x_1,\ldots,x_N) = \det \sum_{a=1}^N A^a_{ij} x_a\,,
\label{Baikov:P}
\end{equation}
and the normalization constant is
\begin{equation*}
C = \frac{\pi^{-L(L-1)/4-LE/2}}{\prod_{i=1}^L \Gamma\left(\frac{d-M+i}{2}\right)}
G(p_1,\ldots,p_E)^{(-d+E+1)/2} \det A^a_{ij}
\end{equation*}
(in the last determinant, the pair $(i,j)$ with $j\ge i$
is considered as a single index).

Acting by the operator $\mathbf{a}^-$ on~(\ref{Baikov:Baikov})
multiplies the integrand by $x_a$;
acting by $\hat{\mathbf{a}}^+$ replaces $1/x_a^{n_a}$
by $n_a/x_a^{n_a+1}=-\partial_a(1/x_a^{n_a})$,
we integrate by parts and take into account the fact
that the polynomial $P$ vanishes at the boundaries:
\begin{equation}
\begin{split}
&\mathbf{a}^- I = C \int \frac{d x_1 \cdots d x_N}{x_1^{n_1} \cdots x_N^{n_N}}
x_a P^{(d-M-1)/2}\,,\\
&\hat{\mathbf{a}}^+ I = C \int \frac{d x_1 \cdots d x_N}{x_1^{n_1} \cdots x_N^{n_N}}
\partial_a  P^{(d-M-1)/2}\,.
\end{split}
\label{Baikov:Oper}
\end{equation}
Note that $[\partial_a,x_b]=\delta_{ab}$,
in accordance with~(\ref{IBP:comm3}).
The IBP relations $\mathbf{O}_{ij}(\hat{\mathbf{a}}^+,\mathbf{a}^-) I = 0$
become~\cite{B:96,B:97}
\begin{equation}
\mathbf{O}_{ij}(\partial_a,x_a) P^{(d-M-1)/2}(x_a-m_a^2) = 0\,.
\label{Baikov:IBP}
\end{equation}

Let's check that $P(x_a)$~(\ref{Baikov:P}) indeed satisfies this requirement.
If $j\le L$ ($q_j=k_j$),
\begin{equation*}
\begin{split}
\mathbf{O}_{ij}(\partial_a,x_a) &= d \delta_{ij}
- \sum_{a=1}^N \sum_{b=1}^N \sum_{m=1}^M A_a^{mi} A^b_{mj} (1+\delta_{mi})
\partial_a (x_b-m_b^2)\\
&= (d-M-1) \delta_{ij}
- \sum_{a=1}^N \sum_{b=1}^N \sum_{m=1}^M A_a^{mi} A^b_{mj} (1+\delta_{mi})
(x_b-m_b^2) \partial_a
\end{split}
\end{equation*}
(see~(\ref{IBP:k}), note the order of operators).
We substitute
\begin{equation*}
\sum_{a=1}^N A_a^{mi} \partial_a = \partial_{mi} = \frac{\partial}{\partial s_{mi}}\,,\qquad
\sum_{b=1}^N A^b_{mj} (x_b-m_b^2) = s_{mj}\,,
\end{equation*}
and get
\begin{equation*}
\mathbf{O}_{ij} = (d-M-1) \delta_{ij}
- \sum_{m=1}^M (1+\delta_{mi}) s_{mj} \partial_{mi}
\end{equation*}
(this form can be directly obtained from~(\ref{IBP:Oij})
if one takes into account that the order of $s_{mj}$ and $\partial_{mi}$
in~(\ref{IBP:Oij}) is interchanged due to integration by parts
with respect to $x_a$ in~(\ref{Baikov:Oper})).
The derivatives of $P=\det s_{ij}$ are proportional to elements
of the inverse matrix $s^{-1}$:
\begin{equation*}
\partial_{ij} P = (2-\delta_{ij}) P\cdot (s^{-1})_{ji}\,.
\end{equation*}
And now we see that~(\ref{Baikov:IBP}) is satisfied.
The proof at $j>L$ is similar, the only difference is
that $s_{ij}$ with $i,j>L$ (external kinematic quantities)
are not expressed via $x_a$ but kept intact.

The Feynman integral~(\ref{Intro:I}) can be expressed as
\begin{equation}
I(\vec{n}) = \sum_k c_k(\vec{n}) I(\vec{n}_k)\,,\qquad
c_k(\vec{n}_l)=\delta_{kl}\,,
\label{Baikov:masters}
\end{equation}
where $\vec{n}_k$ are the indices of the master integrals.
The coefficients $c_k(\vec{n})$ obey the same IBP relations
\begin{equation}
\mathbf{O}_{ij}(\hat{\mathbf{a}}^+,\mathbf{a}^-) c_k(\vec{n}) = 0
\label{Baikov:IBPc}
\end{equation}
as the integral $I(\vec{n})$, but with different boundary conditions.
Let's consider a sector $n_{1,\ldots,m}>0$, $n_{m+1,\ldots,N}\le0$
(any sector has this form after a suitable re-numbering of $n$'s).
Suppose there is 1 master integral in this sector:
its corner $n_{1,\ldots,m}=1$, $n_{m+1,\ldots,N}=0$.
Then only integrals from this sector and higher ones
can contain this master integral $I(\vec{n}_k)$.
In other words, the coefficient $c_k(\vec{n})$ vanishes
if any index $n_{1,\ldots,m}$ is $\le0$.

Let's consider~\cite{B:96,B:97}
\begin{equation}
\oint \frac{d x_1}{x_1^{n_1}} \cdots \oint \frac{d x_m}{x_m^{n_m}}
\int \frac{d x_{m+1}}{x_{m+1}^{n_{m+1}}} \cdots \int \frac{d x_N}{x_N^{n_N}}
P^{(d-M-1)/2}\,,
\label{Baikov:sol}
\end{equation}
where the contours in the first $m$ integrals are small circles around the origin.
This integral satisfies the IBP relations due to~(\ref{Baikov:IBP})
(because boundary terms from integration by parts in $x_a$ vanish),
and vanishes when any index $n_{1,\ldots,m}$ is $\le0$.
It is natural to assume that it is a linear combination of $c_k(\vec{n})$
and $c_{k'}(\vec{n})$ for master integrals in lower sectors
(they also vanish when any of the indices $n_{1,\ldots,m}$ is $\le0$).
In other words, $c_k(\vec{n})$ is a linear combination of~(\ref{Baikov:sol})
and similar integrals where some more $\int$ are replaced by $\oint$.
Coefficients in this linear combination are fixed
by the boundary conditions~(\ref{Baikov:masters}).

As a simplest example, let's consider the 1-loop massive vacuum diagram (Fig.~\ref{F:v1})
(with $m=1$).
In Euclidean space, $D=k^2+1$.
The Baikov polynomial is $P(x)=x$, and
\begin{equation}
V(n) = C \int_1^\infty \frac{d x}{x^n} (x-1)^{(d-2)/2}
\label{Baikov:v1}
\end{equation}
(it is easy to check that this formula reproduces the standard result for $V(n)$).
There is 1 master integral V(1): $V(n)=c(n)V(1)$.
The coefficient $c(n)$ is
\begin{equation}
\begin{split}
c(n) &\sim \oint \frac{d x}{x^n} (x-1)^{(d-2)/2} \sim
\frac{1}{(n-1)!}
\left.\left(\frac{d}{d x}\right)^{n-1} (x-1)^{(d-2)/2} \right|_{x=0}\\
&\sim \frac{1}{(n-1)!}
\left(-\frac{d}{2}+1\right) \cdots \left(-\frac{d}{2}+n-1\right)\,.
\end{split}
\label{Baikov:c1}
\end{equation}
Taking into account $c(1)=1$, we reproduce~(\ref{IBP:V1sol}).

Now let's consider the 1-loop massless self-energy (Fig.~\ref{F:p1}).
In Euclidean space, $D_1=k^2$, $D_2=(k+p)^2$ (we set $p^2=1$),
therefore the Baikov polynomial is
\begin{equation}
P(x_1,x_2) = \left|
\begin{array}{cc}
x_1 & \frac{x_2-x_1-1}{2}\\
\frac{x_2-x_1-1}{2} & 1
\end{array}
\right| = x_1 - \frac{(x_2-x_1-1)^2}{4}\,.
\label{Baikov:Pp1}
\end{equation}
There is only 1 non-trivial sector, and 1 master integral $G(1,1)$.
The coefficients of this master integral are
\begin{equation}
\begin{split}
c(n_1,n_2) &\sim \oint \frac{d x_1}{x_1^{n_1}} \oint \frac{d x_2}{x_2^{n_2}}
P(x_1,x_2)^{(d-3)/2}\\
&\sim \frac{1}{(n_1-1)!\,(n_2-1)!}
\left.\left(\frac{d}{d x_1}\right)^{n_1-1} \left(\frac{d}{d x_2}\right)^{n_2-1}
P(x_1,x_2)^{(d-3)/2} \right|_{x_1=x_2=0}\,.
\end{split}
\label{Baikov:c2}
\end{equation}

Application of this formalism to reduction of 3-loop vacuum diagrams
is discussed in detail in~\cite{BS:98}.
It was used in the formidable task of reduction of 4-loop massless self-energy
diagrams~\cite{BC:10}.
Some additional examples can be found in~\cite{SS:03} and Chapter~6 of~\cite{FIC}.

Using this formalism, P.\,A.~Baikov~\cite{B:00} has constructed an elegant proof
that the 3-loop massless non-planar integral (the last one in~(\ref{p2:H3b}))
cannot be reduced to lower sectors.
Indeed, suppose such a reduction exists, i.\,e. the equation
\begin{equation}
\raisebox{-6.0mm}{\includegraphics{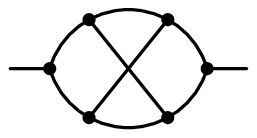}}
- c_1 \raisebox{-6.0mm}{\includegraphics{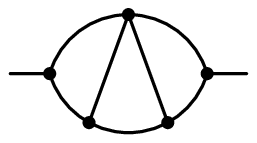}}
- c_2 \raisebox{-6.0mm}{\includegraphics{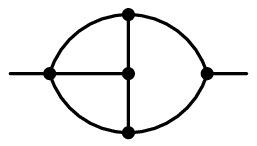}}
- \cdots = 0
\label{Baikov:nonplan}
\end{equation}
is a linear combination of the IBP identities.
Then it must hold not only for the Feynman integrals $I(n_1,\ldots,n_8,n_9)$,
but also for any solution of the IBP relations
(the indices $n_{1\ldots8}$ correspond to 8 denominators,
and $n_9$ to the irreducible numerator).
Let's apply it to
\begin{equation*}
s(n_1,\ldots,n_8,n_9) = \oint \frac{d x_1}{x_1^{n_1}} \cdots \oint \frac{d x_8}{x_8^{n_8}}
\int \frac{d x_9}{x_9^{n_9}} P(x_1,\ldots,x_8,x_9)^{(d-5)/2}\,.
\end{equation*}
This integral vanishes if any of the indices $n_{1\ldots8}$ is $\le0$
(all terms in~(\ref{Baikov:nonplan}) except the first one are absent).
For the first term we have
\begin{equation*}
s(1,\ldots,1,n_9) \sim \int \frac{d x_9}{x_9^{n_9}}
P(0,\ldots,0,x_9)^{(d-5)/2}\,,
\end{equation*}
where $P(0,\ldots,0,x_9)\sim x_9^2(1-x_9)^2$
(for a suitable choice of the irreducible numerator $D_9$~\cite{B:00}).
This means that we may choose the integration contour in $x_9$
to go from 0 to 1 (so that there are no boundary terms,
and $s$ satisfies the IBP relations).
Then $s(1,\ldots,1,0)\ne0$, and we have a contradiction.
This means that~(\ref{Baikov:nonplan}) cannot follow from the IBP relations.

Why doesn't a similar reasoning apply to the planar diagram
(the second one in~(\ref{p2:H3t})) which is reducible?
For this diagram, $P(0,\ldots,0,x_9)\sim x_9^2$,
and we cannot choose a suitable integration contour for $x_9$.

A form of the irreducibility criterion suitable for application
in complicated problems (such as 4-loop massless self-energies)
has been proposed in~\cite{B:06}.

\section{Conclusion}
\label{S:Conc}

Approaches to the problem of reduction of Feynman integrals
can be classified as following:
\begin{itemize}
\item Generic $n_i$
\begin{itemize}
\item Construct an algorithm and implement by hand: \texttt{Mincer}, \dots
\item More automated approaches
\begin{itemize}
\item Gr\"obner-bases methods
\item Lie-algebra based method
\item Baikov's method
\end{itemize}
\end{itemize}
\item Specific numeric $n_i$: Laporta algorithm
(\texttt{AIR}, \texttt{FIRE}, \texttt{Reduze}\dots)
\end{itemize}

The most straightforward approach to reduction is to substitute
specific integer values for all the indices $n_a$ and to solve
the resulting huge linear system.
If we consider some region of size $R$ around the origin,
there is $L(L+E)$ relations per point in this region,
and only 1 unknown per point
(plus some unknowns outside the region near its surface,
but their number is proportional to the area of the surface,
and becomes negligible as compared to the volume of the region
at sufficiently large $R$).
The system of IBP relations is highly redundant.
In each sector, we can use these relations to reduce more complicated integrals
to simpler ones (with respect to some total order);
integrals from lower sectors are considered trivial and already known.
A few integrals which cannot be reduced any further are the master integrals.

This is called the Laporta algorithm~\cite{L:00}.
It may require solving huge linear systems.
There are several publicly available implementations:
\texttt{AIR}~\cite{AIR} (in \texttt{Maple}),
\texttt{FIRE}~\cite{FIRE} (in \texttt{\textit{Mathematica}}),
\texttt{Reduze}~\cite{Reduze}
(in \texttt{C++}, using the \texttt{GiNaC} library~\cite{GiNaC}).
There are also numerous implementations which are not publicly available
(several in Karlsruhe, several in Edmonton, by M.~Czakon, etc.;
\texttt{FIRE} version 2 is rewritten in \texttt{C++}
(with \texttt{Fermat}~\cite{Fermat}),
but it is also not public).

Many implementations of the Laporta algorithm are written in \texttt{C++};
for algebraic operations they either use \texttt{GiNaC},
or talk to an external CAS (e.\,g., \texttt{Fermat}).
They have to use very large number of expressions;
in order not to exceed the size of the main memory,
key--value databases (e.\,g., Kyoto Cabinet~\cite{Kyoto}) may be used.

It is possible to reduce the redundancy of the system of IBP relations~\cite{L:08}
using its Lie algebra structure~(\ref{IBP:comm}).
The set of relations
\begin{equation}
\begin{split}
&\partial_i \cdot k_{i+1}\qquad
(i\in[1,L],\;k_{L+1} \equiv k_1)\,;\\
&\partial_1 \cdot p_j\qquad
(j\in[1,E])\,;\\
&\sum_{i=1}^L \partial_i \cdot k_i
\end{split}
\label{Conc:Lee}
\end{equation}
is sufficient: all the other $O_{ij}$ can be obtained by commutators
from this smaller subset.
It is still redundant: there are $L+E+1$ relations per point,
but this is smaller than $L(L+E)$ relations per point
in the complete set.

The Laporta algorithm is universal.
However, it requires one to solve very large linear systems,
if the indices $|n_a|$ are not small,
because the dimensionality of our integer space is large.
Also, if only one $|n_a|\sim R$ is rather large, the algorithm typically requires one
to consider a region of size $\sim R$ in all directions around the origin.
If a special-purpose algorithm for a family of Feynman integrals
can be constructed, then reduction can be done much more efficiently,
and higher $|n_a|$ are attainable.
Also, reduction of integrals with one large $|n_a|\sim R$
does not require to consider a huge number of integrals
with all $|n_a|\sim R$.

Historically, reduction algorithms were constructed by hand,
and implemented in various computer algebra systems (CASs).
The pioneering program of this kind is \texttt{Mincer}~\cite{Mincer:1,Mincer:2}.
It was used (and is still being used) for solving many physical problems.
Some examples of reduction programs for various classes of Feynman integrals
are listed in Table~\ref{T:Prog}.
It is impossible to produce a complete list,
because special-purpose reduction programs for various diagrams
were written and used in very many papers;
the Table just contains typical examples.

\begin{table}[ht]
\caption{Some programs for reducing specific classes of Feynman integrals}
\label{T:Prog}
\begin{center}
\begin{tabular}{|l|l|l|l|}
\hline
Program & Ref. & CAS & Diagrams\\
\hline
\multirow{2}*{\texttt{Mincer}}
& \cite{Mincer:1} & \texttt{SCHOONSCHIP} &
\multirow{3}*{3-loop massless self-energies}\\
& \cite{Mincer:2} & \texttt{FORM}        &\\
\cline{1-3}
\texttt{Slicer} & \cite{Slicer} & \texttt{REDUCE} &\\
\hline
\multirow{2}*{\texttt{Recursor}} &
\multirow{2}*{\cite{B:92}} &
\multirow{2}*{\texttt{REDUCE}} &
2-loop massive on-shell self-energies\\
&&&3-loop massive vacuum diagrams\\
\hline
\texttt{SHELL2} & \cite{SHELL2} &
\multirow{2}*{\texttt{FORM}} &
\multirow{2}*{2-loop massive on-shell self-energies}\\
\texttt{ONSHELL2} & \cite{ONSHELL2} &&\\
\hline
& \cite{A:95} &
\multirow{2}*{\texttt{FORM}} &
\multirow{2}*{3-loop massive vacuum diagrams}\\
\texttt{MATAD} & \cite{MATAD} &&\\
\hline
\texttt{Grinder} & \cite{G:00} & \texttt{REDUCE} &
3-loop HQET self-energies\\
\hline
\texttt{SHELL3} & \cite{SHELL3} & \texttt{FORM} &
3-loop massive on-shell self-energies\\
\hline
& \cite{T:97} & \texttt{FORM} &
\multirow{2}*{2-loop self-energies with arbitrary masses}\\
\texttt{Tarcer} & \cite{Tarcer} & \textit{\texttt{Mathematica}} &\\
\hline
\texttt{Loops} & \cite{Loops} & \texttt{REDUCE} &
2-loop massless self-energies\\
\hline
\end{tabular}
\end{center}
\end{table}

In recent years, the problems being considered became much more complicated
(4-loop vacuum and self-energy diagrams).
They cannot be solved using this traditional approach.
Several attempts to automate construction of IBP reduction algorithms
were made (see Sects.~\ref{S:IBP}, \ref{S:SBases}, \ref{S:Baikov}).
Unfortunately, the problem has not been completely solved.
It seems to be a well-defined mathematical problem,
and I do hope that some universal and elegant solution will appear.

A short comment on CASs used for implementing IBP reduction algorithms is in order.
They form a wide spectrum.
On one side, there is \texttt{\textit{Mathematica}}:
very convenient, with huge amount of built-in mathematical knowledge,
with an advanced GUI; but very hungry with respect to both memory
and CPU time (and also expensive).
On the opposite end of this spectrum there is \texttt{FORM}~\cite{FORM}:
low-level, nearly no built-in mathematical knowledge,
but efficient and suitable for huge calculations.
It has been recently released under GPL.
\texttt{REDUCE}~\cite{REDUCE,G:97} is in the middle:
a convenient high-level language,
a lot of mathematical knowledge (integrals, expansion in series, and much more),
much more efficient than \texttt{\textit{Mathematica}};
it has no fancy GUI.
A few years ago it became free software (BSD license).
If the problem being considered is so large that the size of expressions
is larger than the main memory of the computer,
\texttt{FORM} is the only choice.
It can work with expressions stored on disk efficiently.
If any other system begins to swap, the situation is hopeless.
On the other hand, if the problem fits in the main memory,
\texttt{REDUCE} is a very reasonable candidate.
Its language is well suited for this kind of problems
(I've written a large package \texttt{Grinder} in it).

Several years ago, an interesting benchmark was run
on many different CASs~\cite{Fateman}.
It was multiplication of large sparse multivariate polynomials.
Unsurprisingly, specialized polynomial systems
(\texttt{Pari}~\cite{Pari}, \texttt{Fermat}~\cite{Fermat})
were at the top.
The fastest general-purpose system was \texttt{REDUCE}
(closely followed by \texttt{maxima} compiled with \texttt{CMUCL},
an efficient common lisp compiler).
\texttt{FORM} was 4 times slower than \texttt{REDUCE};
\texttt{Maple} 10 times slower;
and \texttt{\textit{Mathematica}} nearly 20 times slower.
\texttt{REDUCE} has good implementations of polynomial GCD
(important for working with rational functions, and hence for IBP reduction)
and of polynomial factorization, including multivariate
(both of these things are impossible or very difficult in \texttt{FORM}).
\texttt{REDUCE} was the most widely used CAS in physics at the dawn of the CAS era,
but its use has dramatically reduced later;
now, when it's free software, it would be good to use this excellent system more often.
But, as I said, \texttt{FORM} has a unique advantage:
it is the only system that can work with expressions which are larger
than the main memory of the computer.

\begin{sloppypar}
\textbf{Acknowledgements}.
I am grateful to P.\,A.~Baikov, K.\,G.~Chetyrkin, A.\,V.~Kotikov,
R.\,N.~Lee, A.\,V.~Smirnov, and V.\,A.~Smirnov
for reading the manuscript and suggesting improvement;
to D.\,J.~Broadhurst and M.~Steinhauser
for discussions of various aspects of multiloop calculations;
and to S.-O.~Moch
for inviting me to give lectures at CAPP-2011.
This work was supported by the BMBF through grant No.\ 05H09VKE.
\end{sloppypar}

\appendix
\section{Dimensional regularization}
\label{S:DR}

During these lectures, we are going to live in a $d$-dimensional space--time
(1 time and $d-1$ space dimensions, the signature $+-\cdots-$).
The dimensionality $d$ is a symbol;
comparing it to integers ($d>4$ or $d<4$) makes no sense.
If you want to consider a 100-dimensional subspace of the $d$-dimensional space,
this is perfectly legal;
there will also be a $(d-100)$-dimensional subspace orthogonal to your one.

An $L$-loop Feynman integral $I$~(\ref{Intro:I})
has dimensionality of mass to the power $Ld-n$,
where $n$ is an integer dimensionality of the denominators.
If $I$ contains no dimensional parameter,
we can write no result for such an integral except 0.
For example,
\begin{equation}
\raisebox{-7.8mm}{\begin{picture}(18,18)
\put(9,9){\makebox(0,0){\includegraphics{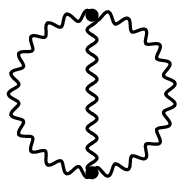}}}
\end{picture}} = 0\,,\qquad\qquad
\raisebox{-5.8mm}{\begin{picture}(28,14)
\put(14,7){\makebox(0,0){\includegraphics{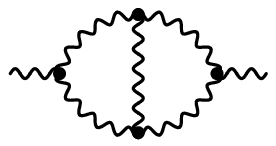}}}
\put(5,4){\makebox(0,0)[r]{$p^2=0$}}
\end{picture}} = 0\,,\qquad\qquad
\raisebox{-5.8mm}{\begin{picture}(28,14)
\put(14,7){\makebox(0,0){\includegraphics{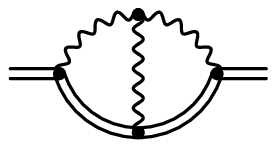}}}
\put(5,4){\makebox(0,0)[r]{$p\cdot v=0$}}
\end{picture}} = 0\,,
\label{DR:0}
\end{equation}
and so on.

If $I$ contains 1 dimensional parameter,
its power is given by dimensions counting:
\begin{equation}
\begin{split}
&\raisebox{-7.8mm}{\begin{picture}(18,18)
\put(9,9){\makebox(0,0){\includegraphics{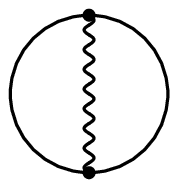}}}
\end{picture}} = m^{Ln-n} I(d)\,,\qquad\qquad
\raisebox{-5.8mm}{\begin{picture}(28,14)
\put(14,7){\makebox(0,0){\includegraphics{p20.eps}}}
\put(4,4){\makebox(0,0)[r]{$p$}}
\end{picture}} = (-p^2)^{Ld/2-n} I(d)\,,\\
&\raisebox{-5.8mm}{\begin{picture}(28,14)
\put(14,7){\makebox(0,0){\includegraphics{h2.eps}}}
\put(4,4){\makebox(0,0)[r]{$p$}}
\end{picture}} = (-2p\cdot v)^{Ld-n} I(d)\,,\qquad\qquad
\raisebox{-5.8mm}{\begin{picture}(28,14)
\put(14,7){\makebox(0,0){\includegraphics{m2.eps}}}
\put(5,4){\makebox(0,0)[r]{$p^2=m^2$}}
\end{picture}} = m^{Ld-n} I(d)\,,
\end{split}
\label{DR:1}
\end{equation}
where $I(d)$ are dimensionless functions of 1 variable $d$.

If $I$ contains 2 dimensional parameters,
it is a non-trivial function of their ratio (in addition to $d$):
\begin{align}
&\raisebox{-5.8mm}{\begin{picture}(28,14)
\put(14,7){\makebox(0,0){\includegraphics{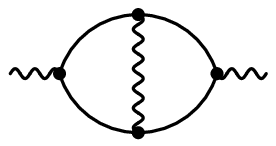}}}
\put(4,4){\makebox(0,0)[r]{$p$}}
\end{picture}} =
m^{Ld-n} I\left(d;\frac{-p^2}{m^2}\right)\,,\qquad\qquad
\raisebox{-5.8mm}{\begin{picture}(28,14)
\put(14,7){\makebox(0,0){\includegraphics{m2.eps}}}
\put(4,4){\makebox(0,0)[r]{$p$}}
\end{picture}} =
m^{Ld-n} I\left(d;\frac{-p^2}{m^2}\right)\,,
\nonumber\\
&\raisebox{-5.8mm}{\begin{picture}(28,15.5)
\put(14,7.75){\makebox(0,0){\includegraphics{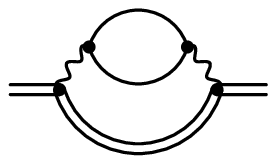}}}
\put(4,4){\makebox(0,0)[r]{$p$}}
\end{picture}} =
m^{Ld-n} I\left(d;\frac{-2p\cdot v}{m}\right)\,,\qquad\qquad
\raisebox{-7.8mm}{\begin{picture}(18,18)
\put(9,9){\makebox(0,0){\includegraphics{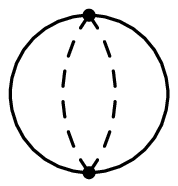}}}
\end{picture}} =
m^{Ld-n} I\left(d;\frac{m'}{m}\right)\,,
\nonumber\\
&{}\hspace{10mm}\raisebox{-5.8mm}{\begin{picture}(28,15.5)
\put(14,7.75){\makebox(0,0){\includegraphics{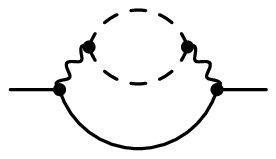}}}
\put(5,4){\makebox(0,0)[r]{$p^2=m^2$}}
\end{picture}} =
m^{Ld-n} I\left(d;\frac{m'}{m}\right)\,.
\label{DR:2}
\end{align}
Integrals with larger numbers of scales non-trivially depend
on all dimensionless ratios.

\section{Gr\"obner bases}
\label{S:Groebner}

Here we briefly introduce Gr\"obner bases for systems of polynomial equations
with commuting variables.
For more detail, see, e.\,g., the textbook~\cite{var}.

\subsection{Statement of the problem}

Suppose we have $n$ variables $x_1$, \dots, $x_n$.
They are not independent, but satisfy some polynomial equations
$p_1=0$, \dots, $p_m=0$ ($p_j$ are polynomials of $x_i$).
Let's consider some polynomial $q$ of the same variables.
It is natural to ask if this polynomial is equal to 0
due to the constraints on our variables or not.
If there is another polynomial $q_2$,
there is the question of their equality.

These questions would become very easy if we had an algorithm
reducing polynomials of dependent variables to a canonical form.
Two equal polynomials reduce to the same canonical form;
a polynomial equal to 0 reduces to the canonical form 0.

We can try to use the equations $p_j=0$
for simplifying the polynomial $q$, i.\,e.\ for replacing
its more complicated terms by combinations of simpler ones.
But to do so we first have to accept some convention
which terms are more complicated and which are more simple.

\subsection{Monomial orders}

We need a total order of monomials
(i.\,e.\ products of powers of the variables $x_1^{n_1}\cdots x_n^{n_n}$).
An order is total if for any monomials $s$ and $t$
either $s<t$, or $s>t$, or $s=t$ is true.
An order is admissible if two properties are satisfied:
\begin{itemize}
\item $1\le s$ for any monomial $s$;
\item if $s<t$ then $su<tu$ for any monomial $u$.
\end{itemize}

One of the most popular monomial orders is lexicographic.
Anybody who has ever seen a dictionary knows it.
We are comparing two monomials:
$s = x_1^{n_1} x_2^{n_2} \cdots x_n^{n_n}$
and $t = x_1^{m_1} x_2^{m_2} \cdots x_n^{m_n}$:
\begin{itemize}
\item $n_1>m_1$ $\Rightarrow$ $s>t$
\item $n_1<m_1$ $\Rightarrow$ $s<t$
\item $n_1=m_1$ $\Rightarrow$
\begin{itemize}
\item $n_2>m_2$ $\Rightarrow$ $s>t$
\item $n_2<m_2$ $\Rightarrow$ $s<t$
\item $n_2=m_2$ $\Rightarrow$\\
\dots
\end{itemize}
\end{itemize}

Another popular order is by total degree than lexicographic.
First we compare the total degree $n=n_1+n_2+\cdots+n_n$ of the monomial $s$
and the total degree $m=m_1+m_2+\cdots+m_n$ of the monomial $t$:
\begin{itemize}
\item $n>m$ $\Rightarrow$ $s>t$
\item $n<m$ $\Rightarrow$ $s<t$
\item $n=m$ $\Rightarrow$ compare lexicographically
\end{itemize}

\subsection{Reduction of polynomials}

Let's fix some admissible monomial order.
We'll write polynomials in descending order:
the leading term first, followed by the rest ones.
We'll normalize all polynomials $p_j$
in such a way that the coefficient of the leading term is 1.
Now they can be used as substitutions which replace the leading term
by minus sum of the remaining ones.
I.\,e. if some term of a polynomial $q$
is divisible by the leading term of some polynomial $p_i$,
we remove this leading term and insert minus sum of the remainder terms of $p_i$ instead.
This is called reduction of the polynomial $q$
with respect to the set of polynomials $p_i$;
if none of the substitutions is applicable,
the polynomial $q$ is called reduced.

\begin{figure}[ht]
\begin{center}
\begin{picture}(60,50)
\put(20,25){\makebox(0,0){$q = x^2 y$}}
\put(35,35){\makebox(0,0){\includegraphics{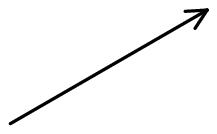}}}
\put(50,45){\makebox(0,0){$q_1 = - y^3 + y$}}
\put(30,37){\makebox(0,0)[r]{$\begin{array}{l}p_1:\\x^2 \to - y^2 + 1\end{array}$}}
\put(35,15){\makebox(0,0){\includegraphics{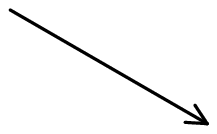}}}
\put(50,5){\makebox(0,0){$q_2 = \tfrac{1}{4} x$}}
\put(30,13){\makebox(0,0)[r]{$\begin{array}{l}p_2:\\x y \to \tfrac{1}{4}\end{array}$}}
\end{picture}
\end{center}
\caption{Reducing $q$ with respect to $p_1$ and $p_2$}
\label{F:p12}
\end{figure}

For example, let's fix the lexicographic order with $x>y$,
and consider the set of polynomials
\begin{equation}
p_1 = x^2 + y^2 - 1\,,\qquad
p_2 = x y - \tfrac{1}{4}\,.
\label{Groebner:p12}
\end{equation}
We want to reduce $q = x^2 y$ with respect to this set.
This can be done in 2 ways:
we can first reduce $q$ with respect to $p_1$ or $p_2$
(Fig.~\ref{F:p12}).

So, we have obtained two different results, $q_1$ and $q_2$;
both cannot be reduced further.
In fact they are equal due to $p_1=0$ and $p_2=0$,
but this is not evident.
Every time when more than one substitution
can be applied to a term of a polynomial $q$
(in this particular case, we can replace either $x^2$ or $x y$ in $x^2 y$),
a fork appears;
maybe, its branches join later, but maybe, they don't
(as in this case).
This example motivates the following definition:

A set of polynomials $p_1$, \dots, $p_n$
is called a \textbf{Gr\"obner basis} (for a given monomial order)
if reduction of any polynomial $q$ with respect to this set is unique.

This definition is not constructive:
it does not say how to check if a given set of polynomials forms a Gr\"obner basis.
There exist equivalent constructive definitions.

\subsection{$S$-polynomials}

In our example, the constraints $p_1=0$ and $p_2=0$
allow us to simplify the monomials $x^2$ and $xy$.
Do these constraints contain an extra information
usable for simplification but not obvious?
Yes, they do!
Let's multiply $p_1$ and $p_2$ by monomials (i.\,e.\ products of powers of variables)
in such a way that their leading terms become identical
(equal to the least common multiple of the leading terms of $p_1$ and $p_2$).
Then we subtract the second polynomial from the first one.
The leading terms cancel, and we get a new polynomial with a new leading term
which can be used for simplifying terms in $q$ (because this new polynomial also vanishes).
This polynomial is called the $S$-polynomial $S(p_1,p_2)$
(from the word subtraction). In our example
\begin{equation*}
\begin{array}{l@{\hspace{10mm}}l@{\hspace{10mm}}l}
p_1 = x^2 + y^2 - 1 = 0
&{}\times y
&x^2 y + y^3 - y = 0\\[2mm]
p_2 = xy - \tfrac{1}{4} = 0
&{}\times x
&x^2 y - \tfrac{1}{4} x = 0\\[2mm]
\cline{3-3}\\
&&\tfrac{1}{4} x + y^3 - y = 0
\end{array}
\end{equation*}
This polynomial can be added to the system of constraints $p_1=0$, $p_2=0$.
Let's normalize its leading coefficient to 1:
\begin{equation}
p_3 = x + 4 y^3 - 4 y\,.
\label{Groebner:p3}
\end{equation}

\begin{figure}[ht]
\begin{center}
\begin{picture}(70,50)
\put(20,25){\makebox(0,0){$q = x^2 y$}}
\put(35,35){\makebox(0,0){\includegraphics{ru.eps}}}
\put(50,45){\makebox(0,0){$q_1 = - y^3 + y$}}
\put(30,37){\makebox(0,0)[r]{$\begin{array}{l}p_1:\\x^2 \to - y^2 + 1\end{array}$}}
\put(35,15){\makebox(0,0){\includegraphics{rd.eps}}}
\put(50,5){\makebox(0,0){$q_2 = \tfrac{1}{4} x$}}
\put(30,13){\makebox(0,0)[r]{$\begin{array}{l}p_2:\\x y \to \tfrac{1}{4}\end{array}$}}
\put(53,25){\makebox(0,0){\includegraphics{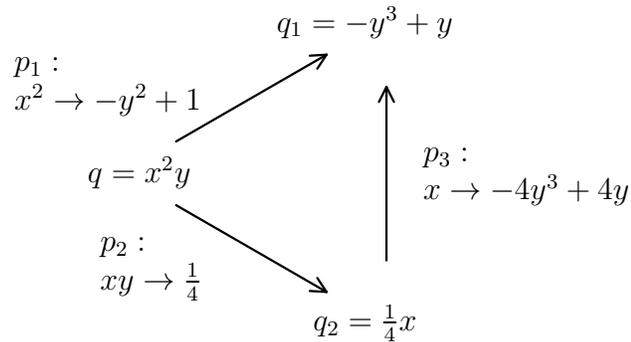}}}
\put(56,25){\makebox(0,0)[l]{$\begin{array}{l}p_3:\\x \to - 4 y^3 + 4 y\end{array}$}}
\end{picture}
\end{center}
\caption{Reducing $q$ with respect to $p_1$, $p_2$, $p_3$}
\label{F:p123}
\end{figure}

Now there is a new possibility for reduction (Fig.~\ref{F:p123}).
Reducing $q_2$ with respect to $p_3$, we again obtain $q_1$.
Reduction of $q$ with respect to the set $p_1$, $p_2$, $p_3$
leads to the unique result.
In fact, $p_1$, $p_2$, $p_3$ form a Gr\"obner basis
(though we have not proved this).

The set of constraints $p_1=0$, $p_2=0$, $p_3=0$
can be simplified by reducing these polynomials with respect to each other
(and omitting identical zeros).
Such a reduced  Gr\"obner basis is
\begin{equation}
P_1 = y^4 - y^2 + \frac{1}{16} = 0\,,\qquad
P_2 = x + 4 y^3 - 4 y = 0\,.
\end{equation}
It has triangular structure:
the first restriction involves only the junior variable $y$;
the second one --- $y$ and $x$.
This is a general feature in the case of lexicographical orders.

\subsection{Buchberger algorithm}

Generalizing this example, we can formulate an algorithm
to construct a Gr\"obner basis from a set of polynomials:
\begin{itemize}
\item Given $P=\{p_j\}$
\item $S=\{\text{the set of pairs \((p_i,p_j)\) of these polynomials with \(i<j\le n\)}\}$
\item \textbf{while} $S$ is not empty
\item\hspace{10mm} choose and remove some pair $(p_i,p_j)$ from $S$;
\item\hspace{10mm} calculate $S$-polynomial $S(p_i,p_j)$;
\item\hspace{10mm} reduce it with respect to $P$;
\item\hspace{10mm} if the result is not 0, add this polynomial to $P$,\\
$\strut$\hspace{10mm} and the corresponding pairs to $S$
\end{itemize}
\noindent
The set of pairs $S$  alternatingly shrinks and grows.
But it can be proved that this process terminates
after a finite number of steps,
and produces a Gr\"obner basis $P$.

\section{Using \texttt{FIRE}}
\label{S:FIRE}

The \texttt{\textit{Mathematica}} package \texttt{FIRE}~\cite{FIRE}
implements the Laporta algorithm and the Gr\"obner-bases method
(Sect.~\ref{S:SBases}).
It can be downloaded from the site~\cite{FIRE}.
You will need 2 files: \texttt{FIRE.tar.gz} and \texttt{SBases\_3.1.0.m}
(this version is newer than the one in the \texttt{tar.gz} file).
Unpack the archive in some directory,
copy \texttt{SBases\_3.1.0.m} to it,
and include this directory into your \texttt{\textit{Mathematica}} \texttt{\$Path}.

Let's investigate the 2-loop massless self-energy diagram (Fig.~\ref{F:Q2}).
The beginning of any program using \texttt{FIRE} is standard:
\begin{verbatim}
<<SBases_3.1.0.m
<<FIRE_3.4.0.m
<<IBP.m
\end{verbatim}
Then we store the lists of the loop and external momenta,
and the list of the propagator denominators $D_a$,
in global variables:
\begin{verbatim}
Internal={k1,k2};
External={p};
Propagators={-(k1+p)^2,-(k2+p)^2,-k1^2,-k2^2,-(k1-k2)^2};
PrepareIBP[]
\end{verbatim}
The set of $D_a$ must be linearly independent,
and all the scalar products must be expressible via $D_a$;
linear denominators are allowed.
There is no special data type for vectors;
\verb|k1^2| and \texttt{k1*k2} are interpreted as $k_1^2$ and $k_1\cdot k_2$.

Let's have a look at the $\partial_1\cdot k_2$ IBP relation
generated by the program:
\begin{verbatim}
IBP[k1,k2]
\end{verbatim}
\begin{align*}
& - a[3] + a[5] - p^2 a[1] Y[1] - a[1] Y[1] Ym[2] - a[1] Y[1] Ym[3] -  a[5] Y[5] Ym[3]\\
&{} - a[3] Y[3] Ym[4] + a[5] Y[5] Ym[4] +  a[1] Y[1] Ym[5] + a[3] Y[3] Ym[5]
\end{align*}
Here $a[3]$ means $\mathbf{n}_3$,
$Y[1]$ means $\mathbf{1}^+$,
and $Ym[2]$ means $\mathbf{2}^-$.
Individual IBP relations may contain scalar products of external momenta only linearly
(here $p^2$; if there were 2 external momenta $p_1$ and $p_2$,
the product $p_1 p_2$ could also appear).
At this stage, their meaning is unambiguous.
However, later, when combining many IBP relations,
you can get, say, $p_1^2 p_2^2$,
and you cannot know if it is $p_1^2 p_2^2$ or $(p_1\cdot p_2)^2$.
Therefore, if is important to substitute some kinematic invariants
for scalar products like $p_1 p_2$ immediately after generating
the IBP relations.
Here we list all the IBP relations:
\begin{verbatim}
startinglist={IBP[k1,k1],IBP[k1,k2],IBP[k1,p],
    IBP[k2,k1],IBP[k2,k2],IBP[k2,p]}/.p^2->-1;
\end{verbatim}
(this list can be shortened to reduce the redundancy, see~(\ref{Conc:Lee})).

The next step is to write down all the symmetries of the diagram.
If the substitution of the line numbers
${1,2,\ldots,N}\to{s_1,s_2,\ldots,s_N}$
is a symmetry, we include $\{s_1,s_2,\ldots,s_N\}$ into the list.
It is not sufficient to provide generators of the symmetry group,
because the program does not try to multiply
the elements of the symmetry group you provide;
all elements, except the identity substitution, have to be included:
\begin{verbatim}
SYMMETRIES={{2,1,4,3,5},
            {3,4,1,2,5},
            {4,3,2,1,5}};
\end{verbatim}

\begin{sloppypar}
Now we are ready to burn the fire:
\begin{verbatim}
Prepare[AutoDetectRestrictions->True]
Burn[]
\end{verbatim}
These lines initialize data structures for using the Laporta algorithm.
The option \texttt{AutoDetectRestrictions->True}
instructs \texttt{FIRE} to detect trivial sectors automatically,
using the Lee's criterion (Sect.~\ref{S:IBP}).
It first appeared in \texttt{SBases} version 3.1.0;
therefore, it is important to download this version.
Alternatively, instead of autodetecting trivial sectors,
you can provide their list by hand:
\begin{verbatim}
RESTRICTIONS={{-1,-1,0,0,0},
              {-1,0,-1,0,0},
              {-1,0,0,0,-1},
              {0,-1,0,-1,0},
              {0,-1,0,0,-1},
              {0,0,-1,-1,0},
              {0,0,-1,0,-1},
              {0,0,0,-1,-1}};
Prepare[]
\end{verbatim}
Each element of the external list designates a set of sectors;
$-1$ in the $a$-th position means sectors with $n_a\le0$;
$+1$ means sectors with $n_a>0$;
$0$ means ``don't care'', all sectors along $n_a$ will do.
Of course, in lists of trivial sectors (\texttt{RESTRICTIONS})
you will never use $+1$.
It is important to include \emph{all} trivial sectors;
but if, say, $n_1\le0$, $n_2\le0$ is sufficient for $I=0$,
you don't have to list $2^3$ sectors with respect to $n_{3,\ldots,5}$,
just write $0$ in these positions.
\end{sloppypar}

The integral $I(n_1,\ldots,n_N)$ is denoted \verb|F[{n1,...,nN}]| in the program
(don't forget that indices corresponding to irreducible numerators are always $\le0$).
The result is computed in terms of the master integrals \verb|G[{n1,...,nN}]|.
Let's calculate
\begin{verbatim}
F[{1,1,1,1,1}]
\end{verbatim}
The program writes some messages explaining what's going on
while applying the Laporta algorithm,
and produces the result
\begin{equation*}
\frac{2 (-10 + 3 d) (-8 + 3 d) G[{0, 1, 1, 0, 1}]}{(-4 + d)^2}
- \frac{2 (-3 + d) G[{1, 1, 1, 1, 0}]}{-4 + d}\,.
\end{equation*}
We see that there are 2 master integrals~(\ref{p2:masters}).
If things go well, you don't need to read messages detailing the progress
of the algorithm; but they may be useful if something goes wrong.
Let's try something more difficult:
\begin{verbatim}
F[{2,2,2,2,2}]
\end{verbatim}
\begin{align*}
& - \bigl(18 (-9 + d) (-7 + d) (-2 + d) (-20 + 3 d) (-16 + 3 d) (-14 + 3 d) (-10 + 3 d) (-8 + 3 d)\\
&\qquad G[{0, 1, 1, 0, 1}]\bigr)/\bigl((-10 + d)^2 (-8 + d)^2 (-4 + d)\bigr)\\
&{} + \frac{4 (-6 + d) (-5 + d) (-3 + d) (-708 + 242 d - 27 d^2 + d^3) G[{1, 1, 1, 1, 0}]}%
{(-10 + d) (-8 + d)}\,.
\end{align*}

\texttt{FIRE} can build $S$-bases in some sectors, and later use them for reduction.
If I add the lines
\begin{verbatim}
BuildBasis[{1, 1, 1, 1, 1}]
BuildBasis[{1, 1, 1, 1, -1}]
BuildBasis[{-1, 1, 1, 1, 1}]
BuildBasis[{-1, 1, 1, -1, 1}]
\end{verbatim}
between \texttt{Prepare} and \texttt{Burn},
then calculation of \verb|F[{2,2,2,2,2}]| takes 2 seconds
instead of 20 (on my computer).

\begin{figure}[ht]
\begin{center}
\begin{picture}(48,48)
\put(24,24){\makebox(0,0){\includegraphics{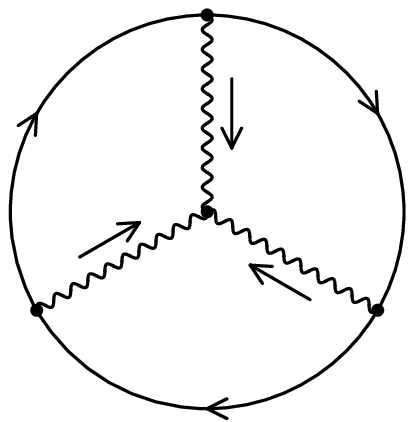}}}
\put(24,1.5){\makebox(0,0){$k_1$}}
\put(4,36){\makebox(0,0){$k_2$}}
\put(44,36){\makebox(0,0){$k_3$}}
\put(27,34){\makebox(0,0)[l]{$k_2-k_3$}}
\put(25,14){\makebox(0,0){$k_3-k_1$}}
\put(12,25){\makebox(0,0){$k_1-k_2$}}
\put(24,6.5){\makebox(0,0){$n_1$}}
\put(11.5,32){\makebox(0,0)[r]{$n_2$}}
\put(42,29){\makebox(0,0)[r]{$n_3$}}
\put(22.5,34){\makebox(0,0)[r]{$n_4$}}
\put(34,21){\makebox(0,0){$n_5$}}
\put(16.5,16){\makebox(0,0){$n_6$}}
\end{picture}
\end{center}
\caption{A 3-loop vacuum diagram}
\label{F:v3}
\end{figure}

Let's consider one more example: 3-loop vacuum diagram~\cite{B:92} in Fig.~\ref{F:v3}.
\begin{verbatim}
<<SBases_3.1.0.m
<<FIRE_3.4.0.m
<<IBP.m
Internal={k1,k2,k3};
External={};
Propagators={1-k1^2,1-k2^2,1-k3^2,-(k2-k3)^2,-(k3-k1)^2,-(k1-k2)^2};
PrepareIBP[]
startinglist={IBP[k1,k1],IBP[k1,k2],IBP[k1,k3],
    IBP[k2,k1],IBP[k2,k2],IBP[k2,k3],
    IBP[k3,k1],IBP[k3,k2],IBP[k3,k3]};
SYMMETRIES={{2,3,1,5,6,4},
            {3,1,2,6,4,5},
            {1,3,2,4,6,5},
            {3,2,1,6,5,4},
            {2,1,3,5,4,6}};
Prepare[AutoDetectRestrictions->True]
Burn[]
F[{1,1,1,1,1,1}]
\end{verbatim}
\begin{equation*}
- \frac{3 (-10 + 3 d) (-8 + 3 d) G[{1, 0, 1, 1, 0, 1}]}{16 (-4 + d) (-7 + 2 d)}
- \frac{(-2 + d)^2 G[{1, 1, 1, 0, 0, 0}]}{8 (-4 + d) (-3 + d)}\,.
\end{equation*}
\begin{verbatim}
F[{2,2,2,2,2,2}]
\end{verbatim}
\begin{align*}
&\bigl(27 (-9 + d) (-6 + d)^2 (-4 + d) (-22 + 3 d) (-20 + 3 d) (-16 + 
      3 d) (-14 + 3 d)\\
&\qquad{} (-10 + 3 d) (-8 + 3 d) (29 - 12 d + d^2) G[{1, 0, 1, 1, 0, 1}]\bigr)\\
&{}/\bigl(16384 (-8 + d) (-7 + d) (-17 + 2 d) (-15 + 2 d) (-13 + 2 d) (-11 + 2 d) (-9 + 2 d) (-7 + 2 d)\bigr)\\
&{} - \frac{(-6 + d)^2 (-4 + d)^2 (-2 + d)^2 (62 - 15 d + d^2) G[{1, 1, 1, 0, 0, 0}]}%
{512 (-8 + d) (-7 + d)^2 (-5 + d) (-3 + d)}\,.
\end{align*}
There are 2 master integrals~(\ref{Equiv:VM}).

If I add the lines
\begin{verbatim}
BuildBasis[{1,1,1,1,1,1}]
BuildBasis[{1,1,1,-1,1,1}]
BuildBasis[{-1,1,1,1,1,1}]
BuildBasis[{1,1,1,-1,-1,1}]
BuildBasis[{-1,1,1,-1,1,1}]
BuildBasis[{1,1,1,-1,-1,-1}]
\end{verbatim}
between \texttt{Prepare} and \texttt{Burn},
calculation of \verb|F[{2,2,2,2,2,2}]| takes 30 seconds
instead of 100.

\end{document}